\documentclass[a4paper,12pt]{article}

\usepackage{amsmath,amssymb,amsfonts} 
\usepackage{graphicx}                 
\usepackage{color}                    
\usepackage{hyperref}                 
\usepackage{cite}

\definecolor{red}{rgb}{1,0,0}           
\definecolor{green}{rgb}{0,1,0}
\definecolor{blue}{rgb}{0,0,1}
\definecolor{darkblue}{rgb}{0,0,0.5}
\definecolor{lightblue}{rgb}{.5,.5,1}
\definecolor{lightgray}{gray}{.87}          
\definecolor{Dark}{gray}{.20}
\definecolor{pink}{rgb}{.95,0.82,0.92}  
\definecolor{yellow}{rgb}{1,1,0}
\definecolor{lightyellow}{rgb}{1,1,.5}
\definecolor{purple}{rgb}{0.7,0,0.85}
\definecolor{darkgreen}{rgb}{0,0.5,0}
\definecolor{orange}{rgb}{0.8,0.2,0.2}
\def \be {\begin{equation}}
\def \ee {\end{equation}}
\def \bea {\begin{eqnarray}}
\def \eea {\end{eqnarray}}
\def \nn {\nonumber}

\def \rr {\raise.35ex\hbox{\small $\prime$}\kern-.17em{\mbox{\large $\imath$}}}
\def \del {\partial}
\def \dels {\partial\kern-.5em / \kern.5em}
\def \As {{A\kern-.5em / \kern.5em}}
\def \Ds {D\kern-.7em / \kern.5em}

\def \a {\alpha}

\def \b {\beta}

\def \om {\omega}

\def \th {\theta}

\def \II {I\hspace{-.1em}I\hspace{.1em}}
\def \III {I\hspace{-.1em}I\hspace{-.15em}I\hspace{.1em}}

\newcommand{\detail}[1]{}

\newcommand{\hide}[1]{}

\begin{document}

\pagestyle{plain}

\begin{titlepage}

\begin{center}

\noindent
\textbf{\LARGE
Static Black Holes With  \\
Back Reaction From Vacuum Energy \\
}
\vskip .5in
{\large 
Pei-Ming Ho
\footnote{e-mail address: pmho@phys.ntu.edu.tw},
Yoshinori Matsuo%
\footnote{e-mail address: matsuo@phys.ntu.edu.tw}
}
\\
{\vskip 10mm \sl
Department of Physics and Center for Theoretical Physics, \\
National Taiwan University, Taipei 106, Taiwan,
R.O.C. 
}\\

\vskip 3mm
\vspace{60pt}
\begin{abstract}

We study spherically symmetric static solutions
to the semi-classical Einstein equation 
sourced by the vacuum energy of quantum fields
in the curved space-time of the same solution.
We found solutions that are small deformations of 
the Schwarzschild metric for distant observers,
but without horizon.
Instead of being a robust feature of objects with high densities,
the horizon is sensitive to the energy-momentum tensor
in the near-horizon region.

\end{abstract}
\end{center}
\end{titlepage}

\tableofcontents

\setcounter{page}{1}
\setcounter{footnote}{0}
\setcounter{section}{0}


\section{Introduction}

Since it was discovered that a black hole finally evaporates by the Hawking radiation, 
the information loss paradox has been a longstanding problem in black hole physics. 
The event horizon plays an important role in this problem, 
and it is assumed in many studies on the information loss paradox 
that the event horizon still exists even when the quantum effects are taken into account. 
On the other hand, there are many arguments from the viewpoint of string theory,
most noticeably via the AdS/CFT duality,
that the information cannot be lost by the black hole evaporation. 
There are also many studies that argue the absence of the horizon
\cite{Gerlach:1976ji,FuzzBall,FuzzBall2,Barcelo:2007yk,Vachaspati:2006ki,Krueger:2008nq,Kawai:2013mda,Kawai:2014afa,Ho:2015fja,Kawai:2015uya,Ho:2015vga,Ho:2016acf,Kawai:2017txu,Fayos:2011zza,Mersini-Houghton,Saini:2015dea,Baccetti}. 

In this paper, we will explore the connection between
the near-horizon geometry and the energy-momentum tensor.
We study the back reaction from 
the vacuum energy-momentum tensor of quantum fields on the near-horizon geometry, 
and consider how the vacuum energy-momentum tensor modifies the geometry. 
The motivation comes from the studies with self-consistent treatment of the Hawking radiation 
in geometries of the black hole evaporation. 
Recently it was shown that no horizon forms during the gravitational collapse 
if the back reaction from the Hawking radiation is taken into account 
\cite{Kawai:2013mda,Kawai:2014afa,Ho:2015fja,Kawai:2015uya,Ho:2015vga,Ho:2016acf,Kawai:2017txu}. 

In the study of black holes,
Hawking radiation is associated with a conserved energy-momentum tensor, 
which can be computed as the vacuum expectation value of
the energy-momentum operator of quantum fields outside the horizon.
Naively, 
this quantum correction to the energy-momentum tensor,
being extremely small,
should have very little effect on the black-hole horizon,
which exists at a macroscopic scale.
On the other hand,
the formation of horizons in gravitational collapses
is known to be a critical phenomenon \cite{CriticalPhenomena}.
Infinitesimal modifications to the initial condition around the critical value
can make a significant difference in the final states.
Indeed,
we will show that in some sense
the existence of horizon is very sensitive to 
the variation of the energy-momentum tensor.

As a first step,
we will focus on static configurations with spherical symmetry in this work,
and leave its generalization to dynamical processes without spherical symmetry to the future.
We will demonstrate in two different models of quantum fields that 
the quantum correction to the energy-momentum tensor 
is capable of removing the horizon.

We are not claiming that an infinitesimal modification to the energy-momentum tensor
leads to dramatic changes in physics.
The quantum energy-momentum tensor outside a static star
is extremely weak for a distant observer.
Their back reaction to the geometry can indeed be neglected as a good approximation
for the space-time region outside the horizon which is visible to a distant observer.
On the other hand,
the horizon can be deformed into a wormhole-like geometry
by merely modifying the geometry within an extremely small region near the Schwarzschild radius,
and the difference can be hard to distinguish for a distant observer.

The vacuum expectation value of the energy-momentum operator
has been calculated in the fixed Schwarzschild background
for the models that we will consider,
as well as for other similar models,
but its back reaction to the geometry have been ignored,
or treated with insufficient rigor most of the time.
The fact that the vacuum energy-momentum tensor is consistently small
outside a black hole was taken by many as a confirmation that
its back reaction to the background geometry through 
the semi-classical Einstein equation 
\be
G_{\mu\nu} = \kappa \langle T_{\mu\nu} \rangle
\ee
can be ignored.
However, it is also assumed by some that 
the Boulware vacuum is unphysical
as it has divergence at the horizon 
in the Schwarzschild geometry. 
A circular logic is sometimes used to further argue that 
the back reaction of the quantum effects can be ignored,
since those states with large quantum effects such as 
the Boulware vacuum are all assumed to be unphysical.
However, it is also an unnatural condition 
to introduce the incoming energy such that 
the energy-momentum tensor does not diverge at the future horizon,
unless the future horizon is already proven to exist.
Here, we impose the more natural initial condition that 
there is no incoming energy flow in the past infinity. 
If the Boulware vacuum is unphysical, 
there must be outgoing energy in future infinity and 
black holes cannot have static states for this initial condition. 
However, there is a chance to have a physical state 
for the Boulware vacuum if we take the back reaction from the quantum effects into account. 
We will show nonperturbatively that 
there is a solution to the semi-classical Einstein equation 
for the Boulware vacuum without divergence in the energy-momentum tensor,
and hence, 
it is physically sensible to consider the Boulware vacuum.

The perturbation expansion for the semi-classical Einstein equation 
around the Schwarzschild background breaks down at the horizon.
Due to the divergence in the Boulware vacuum,
the correction term to the Schwarzschild solution also diverges at the horizon.
Instead of the perturbation theory as an expansion in the Newton constant,
we rely on non-perturbative analysis of the semi-classical Einstein equations.
Our analysis shows that the horizon of the classical Schwarzschild solution 
can be deformed into a wormhole-like structure (without horizon)
by an arbitrarily small correction to the energy-momentum tensor.
The wormhole-like structure connects the internal region of the star to
the external region well approximated by the Schwarzschild solution.
We emphasize that the wormhole-like geometry is not connected to another open space
(hence it is not a genuine wormhole),
but to the surface of a matter sphere.
We will not consider the geometry inside the matter sphere,
where the energy-momentum tensor of the matter 
needs to be specified. 
Instead,
we will focus on the neighborhood of the wormhole-like geometry,
or other kinds of geometry
that replaces the near-horizon region.
In the literature,
the wormhole-like geometry is also called
a ``bounce'' or ``turning point'' (of the radius function $r$). 

For static configurations with spherical symmetry,
the event horizon is also a Killing horizon and an apparent horizon. 
An object falling through the horizon can never return.
When the horizon is deformed into a wormhole-like structure,
an object falling towards the center can always return,
but only after an extremely long time.
Hence,
from the viewpoint of a distant observer,
an ``approximate horizon'' still exists.
In practice,
an extremely long period of time beyond a certain infrared cutoff
can be approximated as infinite time.
The horizon can be viewed as the ideal limit 
in which the time for an object to come out of the approximate horizon
approaches to infinity.
In this sense,
our conclusion that
an infinitesimal modification can replace a mathematical horizon 
by an approximate horizon is nothing dramatic.
Nevertheless,
while the notion of horizon plays a crucial role
in conceptual problems such as the information loss paradox,
it is of crucial importance to understand
how to characterize the geometry of approximate horizons
and their difference from the exact horizon.

It should be noted, however, that 
the Killing horizon in static geometry is not 
directly related to the information loss paradox. 
This paper is aimed at exploring 
the local structure around the horizon 
and study how it is modified by quantum corrections, 
and the global structure is out of the scope of this paper. 
We will show that the Killing horizon is sometimes removed
after taking into account the back reaction of the quantum effects. 
This does not immediately imply that the event horizon does not appear 
in the dynamical process of a gravitational collapse, 
as the notion of the event horizon for dynamical systems is 
quite different from that of static systems,
and the horizon might be recovered due to the effect of the Hawking radiation. 
Therefore it is non-trivial to apply the result of this paper 
to the formation of black holes,
which is a problem we will attack in the near future.

After setting up the basic formulation for latter discussions in Sec.~\ref{s-wave},
we revisit in Sec.~\ref{ToyModel} and Sec.~\ref{dilaton-coupled-static}
different models people have used to estimate 
the vacuum expectation value of the energy-momentum operator outside a black hole,
as examples of how tiny quantum corrections can turn off the horizon.
It is not of our concern whether these models are accurate.
Our intention is to demonstrate the possibility for
a small correction in the energy-momentum tensor to remove the horizon. 

In Sec.~\ref{General},
we consider generic static configurations
with spherical symmetry,
without assumptions on the underlying physics
that determines the vacuum energy-momentum tensor.
In addition to Einstein equations,
we only assume that the geometry is free of singularity at macroscopic scales.
(The possibility of a singularity at the origin is expected to be resolved 
by a UV-complete theory 
and is irrelevant to the low-energy physics for macroscopic phenomena.)
It turns out that this regularity condition leads to clear 
connections between the horizon and the energy-momentum tensor at the horizon.
This provides us with a context in which the results of earlier sections 
can be understood.

\section{4D Einstein Equation in S-Wave Approximation}
\label{s-wave}

In this paper,
we assume the validity of
the 4-dimensional semi-classical Einstein equation,
\be
G^{(4)}_{\mu\nu} = \kappa \langle T^{(4)}_{\mu\nu} \rangle \ ,
\label{Einstein-Eq}
\ee
in which gravity is treated classically but 
the quantum effect on
the energy-momentum tensor is taken into account. 
Assuming that the classical energy-momentum tensor vanishes
outside the radius $R$ of the star,
the energy-momentum tensor for $r > R$ is completely given by
the expectation value $\langle T^{(4)}_{\mu\nu} \rangle$ 
of the quantum energy-momentum operator.

To determine the energy-momentum tensor $\langle T^{(4)}_{\mu\nu} \rangle$ outside the star,
we will consider massless scalar fields as examples
--- except that in Sec.~\ref{General} we will consider a generic energy-momentum tensor.
For simplicity, 
we consider only spherically symmetric configurations,
and separate the angular coordinates $(\th, \phi)$ on the 2-sphere
from the temporal and radial coordinates $(x^0, x^1)$ as 
\be
ds^2 = \sum_{\mu, \nu = 0, \cdots, 3} g_{\mu\nu} dx^{\mu} dx^{\nu} 
= \sum_{\mu, \nu = 0, 1} g^{(2)}_{\mu\nu} dx^{\mu} dx^{\nu} + r^2 d\Omega^2 \ , 
\ee
where $d\Omega^2 = d\theta^2 + \sin^2\theta d\phi^2$ is the metric on the 2-sphere. 
Due to spherical symmetry,
we can integrate out the angular coordinates in the action 
for a 4-dimensional massless scalar field,
and obtain its 2-dimensional effective action as 
\begin{align}
S_m &= 
\frac{1}{2} \int d^4 x \; \sqrt{-g} \sum_{\mu, \nu = 0, \cdots, 3} g^{\mu\nu} \del_{\mu}\chi \del_{\nu}\chi
\nn \\
&= \frac{4\pi}{2} \int d^2 x \; \sqrt{-g^{(2)}} \; r^2 \sum_{\mu, \nu = 0, 1} g^{\mu\nu}_{(2)} \del_{\mu} \chi \del_{\nu} \chi \ .
\label{scalar-action}
\end{align}

Next,
we consider the Einstein-Hilbert action.
The 4-dimensional curvature can be decomposed into 2-dimensional quantities as 
\begin{equation}
 R^{(4)} = R^{(2)} - 6 (\partial \phi)^2 + 4 \nabla^2\phi + 2 \mu^{-2} e^{2\phi} \ ,
\end{equation}
where $R^{(2)}$ is the 2-dimensional scalar curvature
and $\phi \equiv - \log(r/\mu)$ appears as the dilaton field in 2 dimensions.%
\footnote{
We use the same symbol $\phi$ for the dilaton as well as 
the azimuthal angle on the 2-sphere
and hope that this will not lead to any confusion.
}
(The dilaton $\phi$
is originated from the radius $r$ of the integrated 2-sphere,
and $\mu$ is an arbitrary scale parameter.)
After integrating out the angular coordinates, 
the 4-dimensional Einstein-Hilbert action turns into
the 2-dimensional effective action for the dilaton field:
\begin{equation}
 S_\text{EH} = - \frac{1}{16\pi G} \int d^2 x \; \sqrt{-g^{(2)}} \, \mu^2 e^{-2\phi}
\left[ R^{(2)} + 2 (\partial\phi)^2 + 2 \mu^{-2} e^{2\phi} \right] \ . 
\label{S-EH}
\end{equation}
As the 2-dimensional Einstein tensor vanishes identically, 
the equations of motion of
the dimensionally reduced action
only involves the dilaton and a cosmological constant. 

In Secs.~\ref{ToyModel}~and~\ref{dilaton-coupled-static},
we will compute the vacuum energy-momentum tensor $\langle T^{(4)}_{\mu\nu} \rangle$
in different models that have been used in the literature 
on the study of the back reaction of Hawking radiation
(e.g.\cite{Davies:1976ei,Parentani:1994ij,Brout:1995rd,Ayal:1997ab,Barcelo:2007yk}),%
\footnote{%
Charged black holes are also studied using similar approximations 
\cite{Trivedi:1992vh,Strominger:1993yf,Sorkin:2001hf,Hong:2008mw}. 
}
and they have been assumed to capture at least the qualitative features of the problem.%
\footnote{%
Incidentally,
the models for 2D black holes in 
Refs.~\cite{Callan:1992rs,Russo:1992ax,Schoutens:1993hu,Piran:1993tq} 
differ from 4D black holes not only in the matter fields
but also in the gravity action.
}
Those with reservations about the accuracy of these models,
or any other assumption adopted in the calculation below,
should also dismiss the literature based on the same assumptions,
and the implication of this work would be at least this:
The existence of horizon depends on the details of the energy-momentum tensor,
and there is so far no rigorous proof of the presence of horizon
that fully incorporates the back reaction
of the vacuum energy-momentum tensor
in a realistic 4-dimensional theory.

Since 4-dimensional and 2-dimensional energy-momentum tensors are defined by
\begin{align}
 T^{(4)}_{\mu\nu} &= \frac{2}{\sqrt{-g}} \frac{\delta S_m}{\delta g^{\mu\nu}} \ , 
\\
 T^{(2)}_{\mu\nu} &= \frac{2}{\sqrt{-g^{(2)}}} \frac{\delta S_m}{\delta g^{\mu\nu}_{(2)}} \ , 
\end{align}
respectively,
their expectation values are related to each other 
(in the s-wave approximation)
by%
\footnote{
Here we treat the dilaton $\phi$ (or equivalently $r$)
as a classical field
since it is originated from the 4-dimensional classical gravity. 
Only the matter fields are quantized in the semi-classical Einstein equation.
}
\begin{equation}
 \langle T^{(4)}_{\mu\nu}\rangle = \frac{1}{r^2} \langle T^{(2)}_{\mu\nu} \rangle
 \qquad
 (\mu, \nu = 0, 1)
 \label{T4-T2}
\end{equation}
on the reduced 2-dimensional space-time with coordinates $(x^0, x^1)$.
Hence the semi-classical Einstein equation (\ref{Einstein-Eq}) becomes
\begin{equation}
 G^{(4)}_{\mu\nu} = \frac{1}{r^2} \langle T^{(2)}_{\mu\nu}\rangle 
 \qquad
 (\mu, \nu = 0, 1).
\label{Einstein-2D}
\end{equation}
The angular components of the 4-dimensional Einstein equation, 
e.g. $G^{(4)}_{\th\th} = \kappa \langle T_{\th\th}^{(4)} \rangle$,
are equivalent to the equation of motion for the dilaton. 

To avoid potential confusions in the discussion below,
we comment that
the 4-dimensional conservation law for the energy-momentum tensor
\begin{equation}
 \nabla^\mu \langle T_{\mu\nu}^{(4)} \rangle = 0 
 \qquad
 (\mu, \nu = 0, 1, 2, 3)
\label{4D-cons}
\end{equation}
can be expressed in terms of the 2-dimensional tensor $\langle T^{(2)}_{\mu\nu}\rangle$ as 
\begin{equation}
 \nabla^\mu \langle T_{\mu\nu}^{(2)} \rangle - \left(\partial_\mu r^{2}\right) \langle T^{(4)\theta}{}_\theta \rangle = 0
 \qquad
 (\mu, \nu = 0, 1),
\label{Cons-4D-2D}
\end{equation}
which in general violates the naive 2-dimensional conservation law
\be
\nabla^\mu \langle T_{\mu\nu}^{(2)} \rangle = 0
\qquad
(\mu, \nu = 0, 1).
\label{2D-Conserv}
\ee
But if we include the energy-momentum tensor of the dilaton field 
in $T^{(2)}_{\mu\nu}$ together with the matter field,
the last term in (\ref{Cons-4D-2D}) would be cancelled
and the 2-dimensional conservation law (\ref{2D-Conserv}) would hold.

\section{Toy Model: 4D Energy-Momentum From 2D Scalars}
\label{ToyModel}

In this section,
we study the toy model considered by Davies, Fulling and Unruh \cite{Davies:1976ei}
for the vacuum energy-momentum tensor outside a massive sphere.
In this toy model, 
we replace the 4-dimensional scalar field \eqref{scalar-action} by 
the 2-dimensional minimally coupled massless scalar field,
whose action is 
\begin{equation}
 S = \frac{1}{2} \int d^2 x \, \sqrt{-g^{(2)}} 
 \sum_{\mu, \nu = 0, 1} g^{\mu\nu}_{(2)} \del_{\mu} \chi \del_{\nu} \chi \ . 
\label{simple-scalar}
\end{equation}
We shall compute the quantum correction $\langle T_{\mu\nu}^{(2)} \rangle$
to the energy-momentum tensor for this 2-dimensional quantum field theory
and then use eq.\eqref{T4-T2} to estimate 
the 4-dimensional vacuum energy-momentum tensor $\langle T_{\mu\nu}^{(4)} \rangle$.

It should be noted that 
the 2-dimensional minimally coupled scalar \eqref{simple-scalar} 
satisfies the 2-dimensional energy-momentum conservation law (\ref{2D-Conserv}).
Thus,
according to the 4-dimensional conservation law \eqref{Cons-4D-2D}, 
the angular components of the energy-momentum tensor 
for the 2-dimensional minimal scalar must vanish:
\begin{equation}
 \langle T^{(4)}_{\theta\theta} \rangle = \langle T^{(4)}_{\phi\phi} \rangle = 0 \ . 
\label{No-angular}
\end{equation}

\subsection{Energy-Momentum From Weyl Anomaly}
\label{DFU-BeforeCollapse}

For minimally coupled scalar fields, 
the quantum effects for the energy-momentum tensor 
is essentially determined by the conformal anomaly and energy-momentum conservation. 
Here we review the work of Davies, Fulling and Unruh \cite{Davies:1976ei},
where they computed the expectation value of the quantum energy-momentum tensor
for the toy model described above.
They did calculation in the fixed Schwarzschild background without back reaction.
We will consider the back reaction of the quantum energy-momentum tensor
after reviewing their work.

Consider a minimally coupled massless scalar 
with the action (\ref{simple-scalar}) for a given 2-dimensional metric.
According to Davies and Fulling \cite{Davies:1976hi}, 
the quantum energy-momentum operator of this 2-dimensional theory
can be regularized to be consistent with energy-momentum conservation,
but it breaks the conformal symmetry. 
The Weyl anomaly is
\begin{equation}
 \langle T^{(2)\,\mu}{}_\mu\rangle = \frac{1}{24\pi} R^{(2)} \ . 
\label{Weyl-anomaly}
\end{equation}
In the conformal gauge,
the metric is specified by a single function $C$ as
\be
ds^2 = - C(u, v) du dv \ ,
\label{2D-metric}
\ee
and the regularized quantum energy-momentum operator
has the expectation value
(for a certain quantum state to be specified below)
\be
\langle T^{(2)}_{\mu\nu} \rangle = \theta_{\mu\nu} + \frac{R^{(2)}}{48\pi} g_{\mu\nu},
\label{T2}
\ee
where 
the 2-dimensional curvature is
\be
R^{(2)} = \frac{4}{C^3}\left(
C\del_u\del_vC - \del_u C \del_v C
\right),
\ee
and
\begin{align}
\theta_{uu} &= - \frac{1}{12\pi} C^{1/2} \del_u^2 C^{-1/2} \ , 
\label{theta-uu} \\
\theta_{vv} &= - \frac{1}{12\pi} C^{1/2} \del_v^2 C^{-1/2} \ ,
\label{theta-vv} \\
\theta_{uv} &= 0 \ .
\label{theta-uv}
\end{align}

The expressions of $\th_{\mu\nu}$ are not given in a covariant form and 
do not transform covariantly under the coordinate transformation
$u \rightarrow u'(u)$, $v \rightarrow v'(v)$ 
(which preserves the conformal gauge)
because it is the energy-momentum tensor for a specific vacuum state.
Choosing a different set of coordinates $(u, v)$ gives 
the energy-momentum tensor for a different state.  
The vacuum state with
the energy-momentum tensor \eqref{T2}--\eqref{theta-uv} 
is the one with respect to which the creation/annihilation operators in the scalar field
are associated with the positive/negative frequency modes $\{e^{i\omega u}, e^{i\omega v}\}$.

While the trace part of the energy-momentum tensor is fixed by the Weyl anomaly, 
the conservation law implies that
the energy-momentum tensor for any state can always be written in the form
\begin{equation}
 \langle T_{\mu\nu}^{(2)} \rangle = \frac{1}{48\pi} g_{\mu\nu} R^{(2)} + \theta_{\mu\nu} + \widehat T_{\mu\nu} \ .
 \label{ThatT}
\end{equation}
The functions $\widehat T_{\mu\nu}$ are the integration constants arising from
solving the equation of conservation and depend only on $u$ for outgoing modes 
and $v$ for incoming modes.
That is,
\begin{align}
 \widehat T_{uu} &= \widehat T_{uu} (u) \ , 
\\
 \widehat T_{vv} &= \widehat T_{uu} (v) \ , 
\\
 \widehat T_{uv} &= 0 \ , 
\end{align}
namely, $\widehat T_{uu}$ and $\widehat T_{vv}$ are a function of $u$ and that of $v$, respectively. 
The dependence of $\langle T_{\mu\nu}^{(2)} \rangle$ on the choice of states
now resides in $\widehat T_{\mu\nu}$, 
which vanishes for the specific vacuum state associated with the coordinates $(u,v)$ 
in the way described above. 
They can also be fixed by the choice of boundary conditions at the spatial infinity.
The conservation law and Weyl anomaly are preserved
regardless of the choice of these functions.

Now we review the computation by Davies, Fulling and Unruh \cite{Davies:1976ei}
for the quantum energy-momentum tensor outside a 4-dimensional static star
without back reaction.
The 4-dimensional metric for a spherically symmetric configuration can be put in the form
\begin{equation}
 ds^2 = - C du\,dv + r^2 d \Omega^2 \ ,
 \label{metric-C-r}
\end{equation}
with two parametric functions $C(u, v)$ and $r(u, v)$.
Assuming that the star is a massive thin shell of radius $r = R$, 
we have $C = 1$ for the empty space inside the shell ($r < R$)
with the light-cone coordinates denoted by $(U, V)$.
When the back reaction of the vacuum energy-momentum tensor is ignored,
\begin{equation}
 C(r) = 1 - \frac{2M}{r} \ , 
 \label{Schwarzschild-Cr}
\end{equation}
for the Schwarzschild metric outside the shell ($r > R$), 
where $M$ is the mass of the star.
The Schwarzschild radius $a_0$ equals $2M$.

The continuity of the metric at $r = R$ determines
the relation between the coordinate system $(U, V)$ inside the shell
and the coordinate system $(u, v)$ outside the shell as
\be
U = (1-2M/R)^{1/2} u \ , 
\qquad
V = (1-2M/R)^{1/2} v \ .
\ee 
As they are related by a constant scaling factor for a star with constant radius $R$,
the notions about positive/negative frequency modes defined by $(U, V)$ and $(u, v)$ are exactly the same.

The quantum state inside the static mass shell is expected to be the Minkowski vacuum,
for which the positive/negative frequency modes are $\{e^{\pm i \omega U}, e^{\pm i \omega V}\}_{\om > 0}$.
For a large radius $R$,
the density of the shell is small,
and we expect that the quantum state to be continuous across $r = R$.
In other words,
the quantum state just outside the shell at $r = R$
is the vacuum state associated with the positive/negative energy modes
$\{e^{\pm i \omega U}, e^{\pm i \omega V}\}_{\om > 0}$,
or equivalently $\{e^{\pm i \omega u}, e^{\pm i \omega v}\}_{\om > 0}$.

One can use (\ref{T2})--(\ref{theta-uv}) to
compute the energy-momentum tensor for $r > R$
directly with $C$ given by (\ref{Schwarzschild-Cr}).
The results are \cite{Davies:1976ei}
\begin{align}
\langle T^{(2)}_{uu} \rangle &=
\frac{1}{24\pi} \left(\frac{3M^2}{2r^4} - \frac{M}{r^3}\right) \ ,
\label{T2uu}
\\
\langle T^{(2)}_{vv} \rangle &=
\frac{1}{24\pi} \left(\frac{3M^2}{2r^4} - \frac{M}{r^3}\right) \ ,
\label{T2vv}
\\
\langle T^{(2)}_{uv} \rangle &=
\frac{1}{24\pi} \left(\frac{2M^2}{r^4} - \frac{M}{r^3}\right) \ .
\label{T2uv}
\end{align}
This is the energy-momentum tensor for a static star given in Ref.\cite{Davies:1976ei}.
The associated quantum state is called the Boulware vacuum \cite{Boulware}.

The Boulware vacuum has vanishing energy-momentum tensor at $r \rightarrow \infty$.
But the energy-momentum tensor diverges at $r = 2M$
in a generic local orthonormal frame
due to the diverging blue-shift factor at the horizon.
Hence it is conventionally assumed that the radius of the star 
is not allowed to be inside the Schwarzschild radius, 
or equivalently, that the Boulware vacuum is not physical 
if the star is inside the Schwarzschild radius. 
We will see below that,
if the back reaction is taken into consideration,
there is no divergence, 
or very large energy-momentum tensor which induces curvature of the Planckian scale.
The geometry outside a star is perfectly self-consistent and regular, 
even if the star is inside the Schwarzschild radius.
This also implies that the Boulware vacuum is physical 
even for a star inside the Schwarzschild radius, 
but the back reaction must be taken into account. 

\subsection{Turning on Back Reaction}
\label{scalar-static}

Now we turn on the back reaction of the vacuum energy-momentum tensor.
The space-time metric should satisfy the Einstein equation (\ref{Einstein-Eq})
with the vacuum energy-momentum tensor given by (\ref{T4-T2}) and (\ref{T2}).

For a static configuration with spherical symmetry,
the metric can always be written as
\be
 ds^2 = - C(r) dt^2 + \frac{C(r)}{F^2(r)} dr^2 + r^2 d\Omega^2 \ ,
\label{MetricA}
\ee
for some functions $C(r)$ and $F(r)$.
The functions $C(r)$ and $F(r)$ are independent of the time coordinate $t$
due to the time translation symmetry.
The off-diagonal components $dt dr$ are absent 
due to the time-reversal symmetry.
This geometry has the Killing horizon associated to the time-like Killing vector 
$\xi = \partial_t$ at $r=a$ if $C(r=a)=0$. 
The radial coordinate can be redefined from $r$ 
to the tortoise coordinate $r_*$ via
\be
\frac{dr}{dr_*} = F(r) \ ,
\label{tortoise}
\ee
such that the metric is
\be
ds^2 = - C(r) \left[ dt^2 - dr_*^2 \right]
+ r^2(r_\ast) d\Omega^2 \ .
\label{ds2-time-indept}
\ee

We can further define the light-cone coordinates as
\begin{align}
u &= t - r_* \ ,
\\
v &= t + r_* \ ,
\end{align}
and the metric 
\be
ds^2 = - C(v-u) du dv + r^2(v-u) d\Omega^2
\label{metric-C}
\ee
is thus a special case of \eqref{metric-C-r} for some one-variable functions $C(v-u)$ and $r(v-u)$.
Since $r$ is a function of $(v-u)$, 
we can invert the function and view $(v-u)$ as a function of $r$.

For example,
for the Schwarzschild metric,
we have
\begin{align}
C(r) &= 1 - \frac{a_0}{r} \ , 
\label{Schwarzschild-C}
\\
F(r) &= 1 - \frac{a_0}{r}, 
\label{Schwarzschild-F}
\\
r_* &\equiv r + a_0 \log\left(\frac{r}{a_0} - 1\right) \ .
\end{align}

For a static, spherically symmetric configuration,
an apparent horizon is also a Killing horizon.
The reason is as follows.
The apparent horizon is a closed surface on which
outgoing light-like vectors do not expand the area of the surface.
Since the area of a sphere of radius $r$ is $4\pi r^2$ by the definition of the coordinate $r$,
a non-expanding vector must satisfy $dr = 0$,
and for it to be light-like,
we need $ds^2(dr = 0) = 0$.
According to \eqref{MetricA},
this implies that $C(r) = 0$
at some radius $r = a$.
On the other hand,
the Killing horizon is a closed surface on which 
the Killing vector is light-like.
Here the Killing vector refers to the time-translation generator $\del_t$.
It is light-like only if $C(r) = 0$.
Hence we see that $C(r) = 0$ is the condition for both
apparent horizon and Killing horizon.

Plugging the metric (\ref{metric-C}) into the Einstein equation,
the Einstein tensors are 
\begin{align}
G^{(4)}_{uu} &= \frac{2\del_u C\del_u r}{Cr}- \frac{2\del_u^2 r}{r} \ , \\
G^{(4)}_{vv} &= \frac{2\del_v C\del_v r}{Cr}- \frac{2\del_v^2 r}{r} \ , \\
G^{(4)}_{uv} &= \frac{C}{2 r^2} + \frac{2\del_u r\del_v r}{r^2} + \frac{2\del_u\del_v r}{r} \ , 
\\
G^{(4)}_{\theta\theta} &= 
\frac{2 r^2 \left(\partial_u C \partial_v C - C \partial_u \partial_v C\right)}{C^3} 
- \frac{4r \partial_u \partial_v r}{C} \ , 
\end{align}
where $G^{(4)}_{\phi\phi}$ equals $G^{(4)}_{\th\th}$ up to an overall factor of $\sin^2\th$.
By using the relations
\be
\frac{\del r(v-u)}{\del v} = - \frac{\del r(v-u)}{\del u} = \frac{1}{2} F(r) \ ,
\label{def-F}
\ee
which follow \eqref{tortoise},  
the Einstein tensors can be completely expressed 
in terms of the two functions $C(r)$, $F(r)$ as
\begin{align}
G^{(4)}_{uu} &= \frac{F(r)}{2 C(r) r} (F(r)C'(r) - C(r)F'(r)) \ ,
\label{G4uu-1}
\\
G^{(4)}_{vv} &= \frac{F(r)}{2 C(r) r} (F(r)C'(r) - C(r)F'(r)) \ ,
\label{G4vv-1}
\\
G^{(4)}_{uv} &= \frac{1}{2 r^2} (C(r) - F^2(r) - r F(r) F'(r)) \ ,
\label{G4uv-1}
\\
G^{(4)}_{\theta\theta} &= 
- \frac{r^2 F}{2 C^3} \left(F C^{\prime\,2} - F' C C' - F C C''\right) 
+ \frac{r}{C}  FF' \ , 
\label{G4thth-1}
\end{align}
where primes on $C$ and $F$ refer to derivatives with respect to $r$.

Let us now investigate the semi-classical Einstein equation (\ref{Einstein-Eq})
with $\langle T^{(4)}_{\mu\nu} \rangle$ given by eq.\eqref{T4-T2}, 
and $\langle T^{(2)}_{\mu\nu} \rangle$ given by eq.\eqref{T2}--\eqref{theta-uv}
for the Boulware vacuum.
In terms of the functions $C(r)$ and $F(r)$ defined in (\ref{metric-C}) and (\ref{def-F}),
the energy-momentum tensor (\ref{T2})--(\ref{theta-uv}) can be written as
\begin{align}
\langle T^{(2)}_{uu} \rangle &= 
\frac{F(r)}{192\pi C^2(r)} [-3 F(r){C'}^2(r) + 2C(r)(F'(r) C'(r) + F(r) C''(r))] \ ,
\label{T2uu-1}
\\
\langle T^{(2)}_{vv} \rangle &= 
\frac{F(r)}{192\pi C^2(r)} [-3 F(r){C'}^2(r) + 2C(r)(F'(r) C'(r) + F(r) C''(r))] \ ,
\label{T2vv-1}
\\
\langle T^{(2)}_{uv} \rangle &=
\frac{F(r)}{96\pi C^2(r)} [- F(r){C'}^2(r) + C(r)(F'(r) C'(r) + F(r) C''(r))] \ .
\label{T2uv-1}
\end{align}
With the Einstein tensor given in \eqref{G4uu-1}--\eqref{G4uv-1},
the Einstein equations (\ref{Einstein-Eq}) are
(up to an overall factor of $F/(2Cr)$)
\begin{align}
F C' - F' C
- \frac{\alpha}{2} \frac{1}{r} (F' C' + F C'')
+ \frac{3\alpha}{4} \frac{1}{C r} F {C'}^2 = 0 \ ,
\label{CF-1}
\\
\frac{C^2}{F r} - \frac{FC}{r} - F' C
- \frac{\alpha}{2} \frac{1}{r} (F' C' + F C'')
+ \frac{\alpha}{2} \frac{1}{C r} F {C'}^2
= 0 \ ,
\label{CF-2}
\end{align}
where the constant parameter
\be
\alpha = \frac{\kappa N}{24\pi}
\ee
is of the order of the Planck length squared.
The parameter $N$ represents the number of massless scalar fields.

\subsection{Breakdown of Perturbation Theory}
\label{Breakdown-Perturbation}

As the quantum correction to the energy-momentum tensor is extremely small,
one naively expects that the Einstein equations (\ref{CF-1}) and (\ref{CF-2})
can be solved order by order perturbatively in powers of the Newton constant $\kappa$
(or equivalently $\alpha$):
\begin{align}
C(r) &= C_0(r) +\alpha C_1(r) + \alpha^2 C_2(r) + \cdots \ ,
\label{c-expand-0}
\\
F(r) &= F_0(r) + \alpha F_1(r) + \alpha^2 F_2(r) + \cdots \ . 
\end{align}
The leading order terms $C_0$ and $F_0$ 
are expected to be given by the Schwarzschild solution
(see \eqref{Schwarzschild-C} and \eqref{Schwarzschild-F}):
\be
C_0(r) = 1 - \frac{a_0}{r} \ ,
\label{Schwarzschild-C0}
\ee
and
\be
F_0(r) = dr/dr_* = \left(\frac{dr_*}{dr}\right)^{-1} = 1 - \frac{a_0}{r} \ .
\ee

The equations for the first order terms are
\begin{align}
F_0 C'_1 - F'_0 C_1 - C_0 F'_1 + C'_0 F_1 
&= \frac{2\kappa}{r}\langle T^{(2)}_{uu} \rangle_0 ,
\\
\frac{C_1 - 2F_0 F_1}{r} - F_0 F'_1 - F'_0 F_1
&= \frac{2\kappa}{r} \langle T^{(2)}_{uv} \rangle_0 .
\end{align}
Here $\langle T^{(2)}_{\mu\nu} \rangle_0$ are 
given by eqs.(\ref{T2uu})--(\ref{T2uv}) for the Schwarzschild background
as the leading order terms of $\langle T^{(2)}_{\mu\nu} \rangle$ 
in the perturbative expansion.

In the region $r > a_0$,
the equations above can be solved
to obtain the first order correction terms $C_1$ and $F_1$.
However, 
at $r = a_0$,
since $F_0(a_0) = C_0(a_0) = 0$,
these two equations imply
\begin{align}
- \frac{\alpha}{a_0} (C_1 - F_1)
&= \frac{2\kappa}{a_0}\langle T^{(2)}_{uu} \rangle_0\Bigr|_{r=a_0}
= \frac{\alpha}{4 a^3_0} \ ,
\\
\frac{\alpha}{a_0} (C_1 - F_1)
&= \frac{2\kappa}{a_0} \langle T^{(2)}_{uv} \rangle_0\Bigr|_{r=a_0}
= 0 \ ,
\end{align}
unless $C'_1$ or $F'_1$ diverges at $r = a_0$.
Apparently,
these two equations are inconsistent,
and the perturbative expansion fails.
In general,
perturvative expansion breaks down at $r = a_0$
where $C(a_0) = F(a_0) = 0$ if 
\be
C'_0(a_0) = a {F'_0(a_0)}^2 \ .
\label{degeneracy-condition}
\ee

Of course,
as the first order equations are inconsistent only at the point $r = a_0$,
one can solve $C_1$ and $F_1$ for $r > a_0$,
and then define $C_1(a_0)$ and $F_1(a_0)$ by taking the limit $r \to a_0$.
As we will show below,
this leads to divergence in $C_1$ (and $C'_1$) at $r = a_0$,
so that the conclusion remains the same:
the perturbation theory breaks down at the horizon.

Taking the difference of the two Einstein equations (\ref{CF-1}) and (\ref{CF-2}),
we can solve $F(r)$ in terms of $C(r)$:
\be
F(r) = \left[\frac{4C^3(r)}
{4C^2(r) + 4r C(r)C'(r) + \alpha {C'}^2(r)}
\right]^{1/2} \ .
\label{F-from-C}
\ee
Plugging it back into either of the two equations,
we find
\be
 2 r \rho'(r) + (2r^2+\alpha)\rho^{\prime\,2}(r) + \alpha r \rho^{\prime\,3}(r) 
 + (r^2 - \alpha) \rho''(r) = 0 \ ,
\label{f-eq}
\ee
where $\rho(r)$ is defined by
\be
C(r) = e^{2\rho(r)} \ .
\label{rho-def}
\ee
One can check that \eqref{f-eq} is consistent with the assumption 
$\langle T^{(4)}_{\theta\theta}\rangle = 0$,
which can be derived from the Einstein equation 
$G_{\th\th} = \kappa \langle T^{(4)}_{\th\th} \rangle$ using \eqref{F-from-C}. 

Now, 
we consider the perturbative expansion of \eqref{f-eq}. 
We expand $\rho$ as 
\begin{equation}
 \rho(r) = \rho_0(r) + \alpha \rho_1(r) + \cdots \ , 
\end{equation}
which is related to the expansion of $C(r)$ \eqref{c-expand-0} via 
\begin{align}
 C_0 (r) &= e^{2\rho_0(r)} \ , 
& 
 C_1 (r) &= 2 \rho_1(r) C_0(r) \ . 
\end{align}
The solutions of $\rho_0$ and $\rho_1$ to \eqref{f-eq} are
\begin{align}
 \rho_0(r) 
 &= 
 \frac{1}{2} \log c_0 + \frac{1}{2} \log\left(1-\frac{a_0}{r}\right) \ , 
\\
 \rho_1(r) 
 &= 
 - \frac{4 r^2 + a_0^2 + 4 a_0 r (2 c_1 r - 1)}{8 a_0 r^2 (r-a_0)} 
 - \frac{2 r - 3 a_0}{4 a_0^2 (r-a_0)} \log\left(1 - \frac{a_0}{r}\right) \ , 
\end{align}
where $a_0$, $c_0$ and $c_1$ are integration constants. 
The constant $a_0$ is the Schwarzschild radius in the classical limit $\a\to 0$. 
An integration constant in $\rho_1$ is absorbed in $c_0$, 
which is the overall constant of $C(r)$. 
While the divergence in $\rho_0$ at $r\to a_0$ implies $C_0(r)=0$, 
the divergence in $\rho_1$ gives here the divergence in $C_1$. 
Due to the divergence in the higher order terms, 
the perturbative expansion breaks down. 

The divergence in the higher order terms 
is related to that in the vacuum energy-momentum tensor for the Boulware vacuum 
even though the energy-momentum tensor does not diverge in the coordinate system above. 
Though the divergence in the energy-momentum tensor for the Boulware vacuum 
is sometimes considered to imply that the Boulware vacuum is unphysical, 
it just implies the breakdown of the perturbative expansion 
in the semi-classical Einstein equation.

The breakdown of the perturbation theory at $r = a_0$
is not in contradiction with the existence of a solution 
which is well approximated by the classical solution $C_0$ and $F_0$.
We will show that the back reaction is significant only 
within a very small neighborhood ($0 < r - a_0 \ll \alpha/a_0$)
that is extremely close to the Schwarzschild radius.
However,
within this tiny region,
the solution to the semi-classical Einstein equation
cannot be treated perturbatively
in powers of the Newton constant $\kappa$.

\subsection{Non-Perturbative Analysis}
\label{Non-Perturbative-Analysis}

Since the perturbative expansion breaks down around the horizon, 
we have to study the non-perturbative features of eq.\eqref{f-eq}. 
If there is a Killing horizon at $r = a$ 
(it does not have to be equal to the Schwarzschild radius $a_0 = 2M$), 
i.e., if $C(a) = 0$,
we must have $\rho\to-\infty$ at $r = a$,
which in turn implies that $\rho'(r)$ diverges at $r = a$.
Assuming that $\rho'(r)$ diverges at $r = a$ with 
$a\gg\alpha^{1/2}$, 
we must have
\begin{align}
 \rho' \gg \frac{a}{\alpha} \gg \alpha^{-1/2} \ , 
\label{region}
\end{align}
in a region sufficiently close to $r = a$. 
Then the third term, $\alpha r \rho^{\prime\,3}$, dominates in the first 3 terms in (\ref{f-eq}), 
and 
\be
\alpha r \rho^{\prime\,3}(r) + (r^2 - \alpha) \rho''(r) \simeq 0
\label{approx-eq}
\ee
in the limit $r \rightarrow a$.
This equation can be easily solved to give 
the asymptotic solution of $\rho'$ in the limit $r \to a$ 
\be
\rho'(r) \simeq \pm \frac{1}{\sqrt{{\alpha}\log(r^2 - \alpha) + c}} \ ,
\ee
with an integration constant $c$.
The value of $c$ is fixed to be
\be
c = - {\alpha}\log(a^2 - \alpha)
\ee
so that $\rho'$ diverges at $r = a$.
Hence
\be
\rho'(r) \simeq \pm \left[{\alpha}\log\left(\frac{r^2-\alpha}{a^2-\alpha}\right)\right]^{-1/2}
\rightarrow \pm \left(\frac{a^2-\alpha}{2\alpha a}\right)^{1/2} (r-a)^{-1/2}
\label{f-approx}
\ee
as $r \rightarrow a$.
As a result,
\be
C(r) \simeq c_0 e^{2\sqrt{k(r-a)}}
\label{C-approx}
\ee
as $r \rightarrow a$,
where we have chosen the sign in (\ref{f-approx}) 
such that $C(r)$ is an increasing function of $r$,
in view of a smooth continuation of $C(r)$
to the asymptotic region in which the geometry is well approximated by 
the Schwarzschild solution (\ref{Schwarzschild-C0}).
Here $c_0$ is a positive constant and
\be
k \equiv \frac{2(a^2-\alpha)}{\alpha a} \simeq \frac{2a}{\alpha} \ .
\ee

The expression 
(\ref{C-approx}) gives a good approximation 
only when \eqref{region} holds,
that is,%
\footnote{
Using eq.\eqref{F-sol-1} below,
one can show that
a small displacement in $r$ of the order of $\Delta r \sim \alpha/a$
corresponds to a physical length of the order of $\Delta s \sim \alpha^{1/2}$,
which is of the Planck length scale unless $N \gg 1$.
This of course does not imply that we need Planckian physics 
in the region \eqref{domain}
because the curvature is still very small
--- see eq.\eqref{GG-1}.
}
\be
0 \leq r - a \ll \frac{\alpha}{a} \ .
\label{domain}
\ee
As a rough estimate of the complete solution of $C(r)$,
we patch the approximate solution (\ref{C-approx}) 
with (\ref{Schwarzschild-C0}) in the neighborhood where
$r - a \sim {\cal O}(\alpha/a)$.
This determines $c_0$ to be a very small number of order
\be
c_0 \sim {\cal O}\left(\frac{\alpha}{a^2}\right) \ .
\label{c0-order}
\ee
Therefore,
although the value of $C(a)$ is not zero
as it needs
for there to be a horizon,
it is indeed extremely small,
giving a huge blue-shift factor relative to a distant observer.
From the viewpoint of a distant observer,
observations on this geometry will not be very different from 
those on the Schwarzschild geometry,
and we expect that $a \simeq a_0$.

The calculations leading to \eqref{C-approx} serves as a mathematical proof that 
it is impossible for $C(r)$ to vanish anywhere,
and thus there is no horizon.
The quantum correction to the energy-momentum tensor is such that
there is no horizon even if
the radius of the star is much smaller than the classical Schwarzschild radius $a_0 = 2M$.
Due to the back reaction of the quantum energy-momentum tensor,
the property of the Boulware vacuum is dramatically changed,
although the geometry beyond a few Planck lengths outside the Schwarzschild radius
remains well approximated by the Schwarzschild solution.

Let us now describe the geometry that replaces the horizon.
According to (\ref{F-from-C}) and \eqref{C-approx}, 
$F(r)$ behaves as 
\be
F(r) \simeq \sqrt{\frac{4c_0(r-a)}{\alpha k}}
\label{F-sol-1}
\ee
for $r$ sufficiently close to $a$.
In the very small region \eqref{domain},
the metric is approximately given by 
\begin{equation}
 ds^2 \simeq -c_0 dt^2 + \frac{\alpha k\, dr^2}{4(r-a)} + r^2 d\Omega^2 \ . 
\end{equation}
This geometry around $r = a$ resembles that of a wormhole. 
By choosing the origin of the tortoise coordinate such that $r_* = a_*$ when $r=a$, 
we have
\be
r \simeq a + \frac{c_0}{\alpha k} (r_* - a_*)^2 
\ee
as $r \rightarrow a$,
and so the metric is
\be
ds^2 \simeq - [c_0 + {\cal O}(r_* - a_*)] (dt^2 - dr_*^2) + [a^2 + {\cal O}((r_* - a_*)^2)] d\Omega^2 \ .
\ee
It is of the same form as the metric for a static (traversable) wormhole.
In terms of $r_*$,
we can clearly see that the geometry can be smoothly connected to the region $r_*< a_*$, 
although this wormhole-like geometry does not lead to another open space
but merely the interior of a star. 

The wormhole-like geometry of the static star with a radius smaller than the Schwarzschild radius
can therefore be understood in the following way.
With spherical symmetry,
the 3-dimensional space perpendicular to the Killing vector
can be viewed as foliations of 2-spheres with their centers at the origin.
As one moves towards the star from afar,
the surface area of the 2-sphere decreases until reaching a local minimum at $r = a$,
which is the narrowest point of the throat.
There is no singularity at $r = a$,
and the area of the 2-spheres starts to increase beyond this point,
until one reaches the boundary of the star.
After that,
the area of the 2-spheres starts to decrease again,
until the area goes to zero at the origin.

\begin{figure}
\begin{center}
\includegraphics[scale=0.7,bb=0 0 259 170]{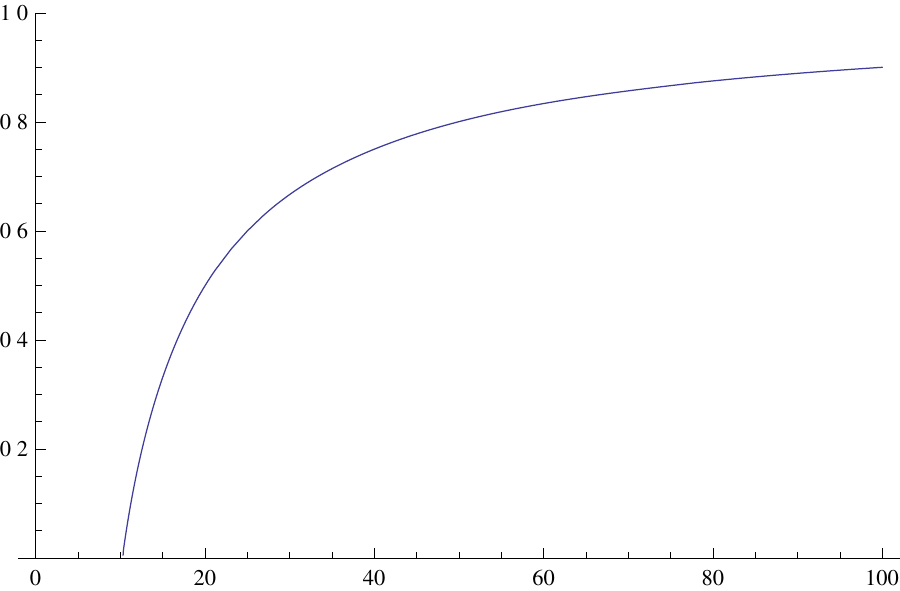}
\hspace{24pt}
\includegraphics[scale=0.7,bb=0 0 259 170]{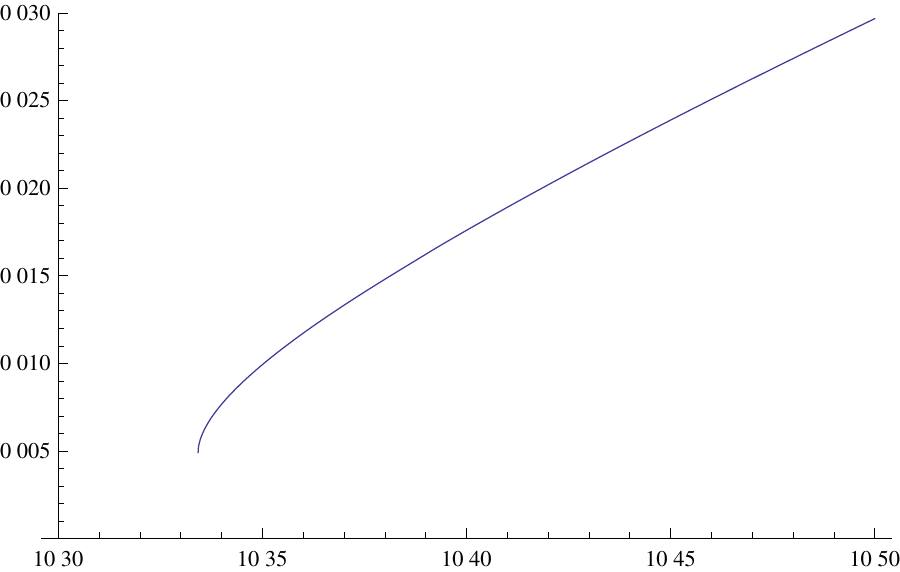}
\caption{\small 
Numerical results for $C$ as a function of $r$. 
$C(r)$ is non-zero (positive) at $r=a$ and well defined only for $r\geq a$.
Left: $C(r)$ vs. $r$ from $r = a$ to $r \gg a$.
Right: $C(r)$ vs. $r$ for a small neighborhood of $r = a$. 
Here, $a_0=10$ and $\alpha=2$. 
}
\label{fig:c}
\end{center}
\end{figure}

\begin{figure}
\begin{center}
\includegraphics[scale=0.7,bb=0 0 259 170]{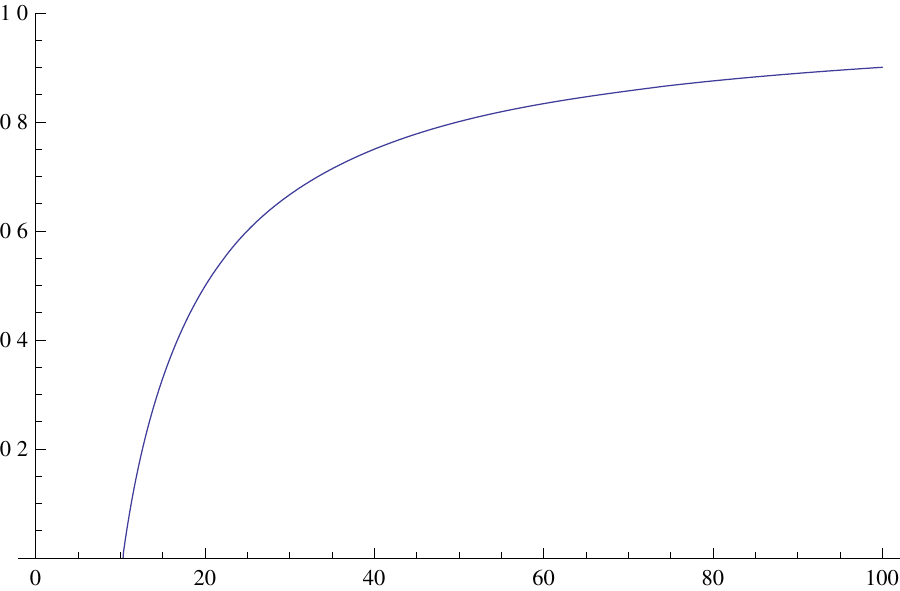}
\hspace{24pt}
\includegraphics[scale=0.7,bb=0 0 259 170]{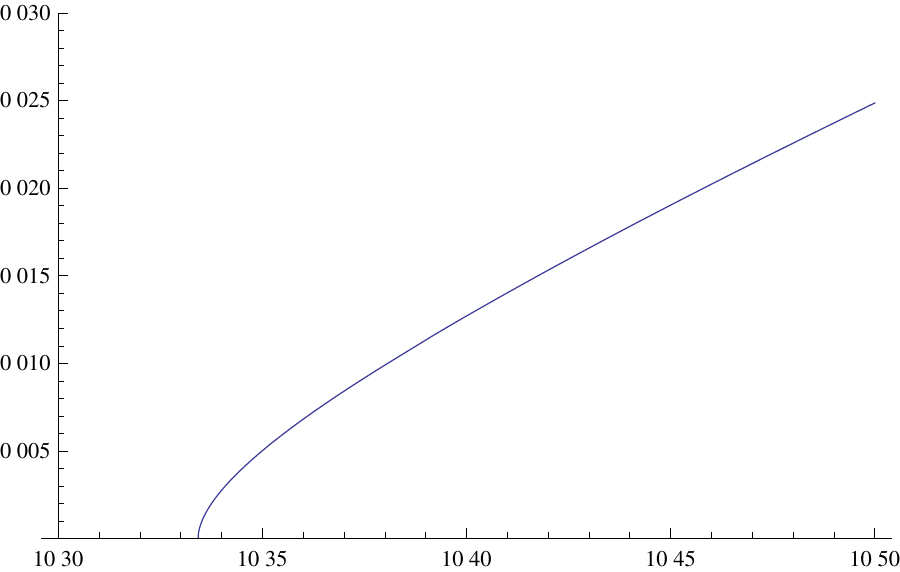}
\caption{\small Numerical results for $F$ as a function of $r$.
$F(r)$ vanishes at $r = a$.
Left: $F(r)$ vs. $r$ from $r = a$ to $r \gg a$.
Right: $F(r)$ in a small neighborhood of $r = a$.
}
\label{fig:f}
\end{center}
\end{figure}

\begin{figure}
\begin{center}
\includegraphics[scale=0.7,bb=0 0 259 170]{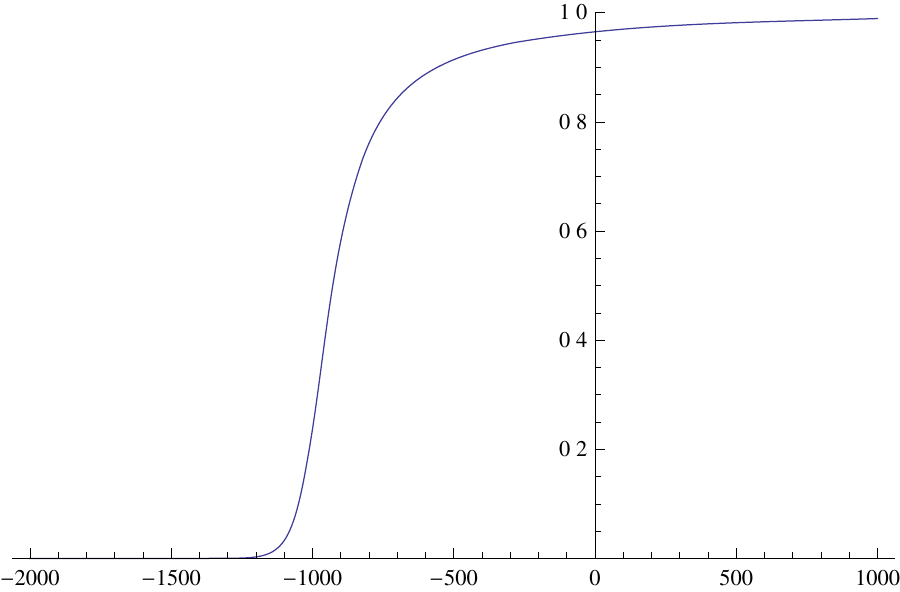}
\\
\includegraphics[scale=0.7,bb=0 0 259 170]{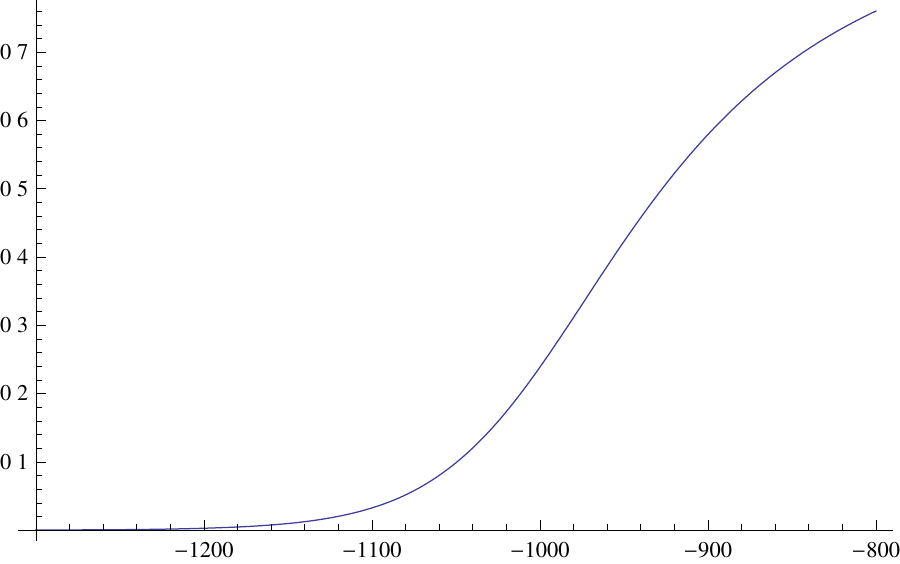}
\hspace{24pt}
\includegraphics[scale=0.7,bb=0 0 259 170]{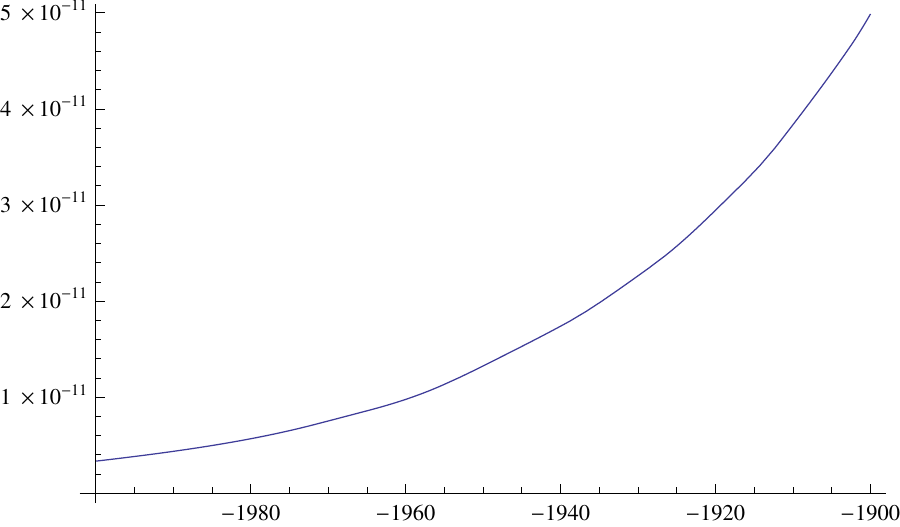}
\caption{\small Numerical results for $C$ as a function of $r_*$. 
Top: $C(r_*)$ vs. $r_*$ from $r_* \ll a_*$ to $r_* \gg a_*$.
Left: $C(r_*)$ vs. $r_*$ in a small neighborhood of $r_*=a_*$.
Right: $C(r_*)$ vs. $r_*$ for $r_* \ll a_*$.}
\label{fig:c-rstar}
\end{center}
\end{figure}

\begin{figure}
\begin{center}
\includegraphics[scale=0.7,bb=0 0 259 170]{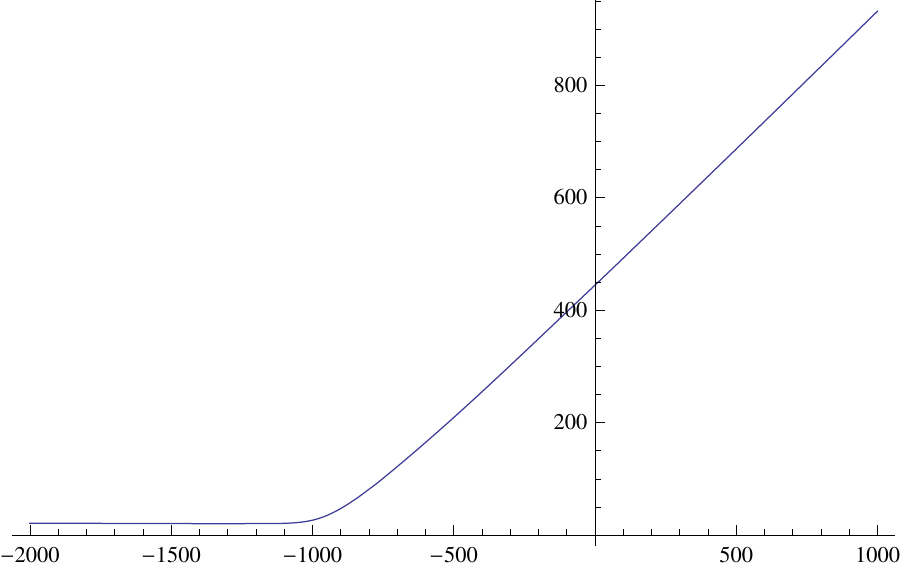}
\hspace{24pt}
\includegraphics[scale=0.7,bb=0 0 259 170]{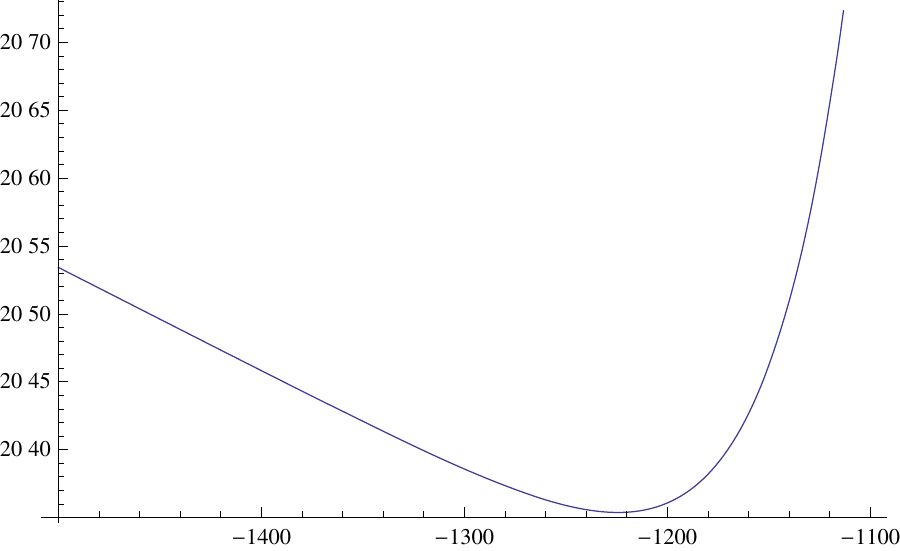}
\caption{\small Numerical results for $r$ as a function of $r_*$.
Left: $r$ vs. $r_*$ from $r_* \ll a_*$ to $r_* \gg a_*$. 
Right: $r$ vs. $r_*$ in a small neighborhood of $r=a$.
The slope is very close to zero but finite for $r_*<a_*$.}
\label{fig:r-rstar}
\end{center}
\end{figure}

In support of our analysis above,
we have solved $C(r)$ and $F(r)$ numerically from eq.\eqref{F-from-C} and \eqref{f-eq},
as shown in Fig.~\ref{fig:c} for $C(r)$ 
and Fig.~\ref{fig:f} for $F(r)$. 
The diagrams for $C(r)$ and $F(r)$ are only plotted for $r \geq a$
simply because $r=a$ is a minimum of $r$.
The numerical simulation for $C$ as a function of $r_*$ is shown in Fig.~\ref{fig:c-rstar},
and the solution can be extended indefinitely in both limits $r_* \to \pm \infty$.
The numerical solution of $r$ as a function of $r_*$ is displayed in Fig.~\ref{fig:r-rstar},
showing that $r$ has a local minimum.

Although the horizon is absent,
i.e. $C(r)$ does not vanish at $r = a$,
the value of $C(a)$ is indeed extremely small 
for a large Schwarzschild radius,
of order ${\cal O}(\alpha/a^2)$ (see \eqref{c0-order}).
The red-shift factor relating the time coordinate $t$ in the neighborhood of $r = a$ 
to the time coordinate $t$ at large $r$
is given by $c_0^{1/2}$.
There is an even larger red-shift for $r < a$.
As a result, 
everything close to or inside the Schwarzschild radius
looks nearly frozen to a distant observer.
For a large Schwarzschild radius,
a real black hole with a horizon and
a wormhole with a large red-shift factor
is very hard to distinguish by observations at distance.

The conventional expectation of the Boulware vacuum is that
the vacuum energy-momentum tensor would diverge at the horizon
if the radius of the star is smaller than the Schwarzschild radius.
But this expectation is based on the calculation that has neglected back reaction.
According to our non-perturbative solution of $C$ and $F$,
in the small neighborhood (\ref{domain}) of $r = a$,
\begin{align}
\langle T^{(2)}_{uu} \rangle &\simeq - \frac{N}{48\pi} \frac{c_0}{\alpha} \sim \mathcal O(1/a^{2}) \ ,
\\
\langle T^{(2)}_{uv} \rangle &\simeq 0 \ ,
\end{align}
and $\langle T^{(2)}_{vv} \rangle$ is the same as $\langle T^{(2)}_{uu} \rangle$.
According to (\ref{c0-order}),
$\langle T^{(2)}_{uu} \rangle$ is of the same order ${\cal O}(1/a^2)$
as its counterpart (\ref{T2uu}) before back reaction is taken into consideration.
$\langle T^{(2)}_{uv} \rangle$ vanishes
as its counterpart does at $r = a_0$.
Since $C(a) = c_0$ is very small \eqref{c0-order},
the energy-momentum tensor at $r=a$ in a local frame is highly blue shifted.
But it is only of order ${\cal O}(\alpha^{-1}a^{-2}))$,
much smaller than the Planck energy density $\alpha^{-2}$.
This invalidates the conventional expectation that
the energy-momentum tensor diverges at the horizon
for the Boulware vacuum.

Since this is no longer a classical vacuum solution,
the Einstein tensor becomes non-zero at $r=a$.
In the small neighborhood (\ref{domain}) around $r = a$,
the Einstein tensor is of order
\be
G^u{}_v \sim G^v{}_u \sim {\cal O}(1/a^2) \ ,
\qquad
G^u{}_u \sim G^v{}_v \sim 0 \ .
\label{GG-1}
\ee
The order of magnitude of $G^u{}_v$
(${\cal O}(1/a^2)$) is small for large $a$,
so that it is consistent to use the low-energy effective description of gravity
(Einstein's equations).

Notice that the disappearance of horizon is not a fine-tuned result.
It is insensitive to many details in eq.\eqref{f-eq},
but only relies on the fact that the dominant terms are $\rho''$ and $\rho'{}^3$.
The appearance of a wormhole-like geometry 
demands that the ratio of the coefficients of these two terms be positive,
but in Sec.~\ref{Tthth=0} below,
we will see that there is still no horizon 
if the ratio is negative,
although the geometry would be different.

We have only considered the local structure at the Schwarzschild radius,
where the near-horizon geometry is replaced by a wormhole-like structure.
It is possible that there is a horizon or singularity deep down the throat.
In fact, 
the result about a wormhole-like structure 
(which was called a ``bounce'')
was first discovered in Ref.\cite{Fabbri:2005nt} via numerical analysis.
In addition they mentioned 
the possibility of a curvature singularity deep down the throat
in the limit $r\to\infty$
(but within finite affine distance) where $C$ goes to zero \cite{Fabbri:2005nt}.
The results of their numerical analysis are completely consistent with 
the discussion in this section,
although our focus is on the near-horizon geometry.
Note that the singularity deep down the throat in the vacuum solution 
is relevant only if the surface of the star does not exist 
until $r\to\infty$ and the mass of the star is localized at the singularity in $r\to\infty$. 
However, 
in a more realistic scenario, 
the surface of the star has a finite area $(r < \infty)$
and so the singularity at $r = \infty$ for the vacuum solution is irrelevant.
The singularity is hence not a robust feature of the wormhole-like geometry.


\subsection{Hartle-Hawking Vacuum}
\label{Hartle-Hawking-Vacuum}

For a more general background, 
the energy-momentum tensor \eqref{ThatT} has the additional terms $\hat{T}_{\mu\nu}$.
For stationary solutions, 
these terms are constants so that
\begin{equation}
 \langle T^{(2)}_{uu} \rangle = \langle T^{(2)}_{vv} \rangle =  
\frac{F(r)}{192\pi C^2(r)} [-3 F(r){C'}^2(r) + 2C(r)(F'(r) C'(r) + F(r) C''(r))] 
+ \frac{b}{48 \pi \a} 
\end{equation}
for some constant $b$.
Then the Einstein equations become
\begin{align}
 F C' - F' C
 - \frac{\alpha}{2} \frac{1}{r} (F' C' + F C'')
 + \frac{3\alpha}{4} \frac{1}{C r} F {C'}^2 - b \frac{C}{r F} &= 0 \ ,
\\
\frac{C^2}{F r} - \frac{FC}{r} - F' C
- \frac{\alpha}{2} \frac{1}{r} (F' C' + F C'')
+ \frac{\alpha}{2} \frac{1}{C r} F {C'}^2
&= 0 \ .
\end{align}
Since the weak energy condition should not be violated in the asymptotic Minkowski space at $r \rightarrow \infty$, 
we shall assume that $b\geq 0$. 
This leads to a positive outgoing energy flux at spatial infinity
as well as an ingoing energy flux of the same magnitude.
The conventional interpretation for this boundary condition is that 
the Hawking radiation from the black hole is balanced
by an ingoing energy flux from a thermal background at the Hawking temperature,
and the corresponding quantum state is called the Hartle-Hawking vacuum.

Due to the energy flux at spatial infinity,
the asymptotic geometry at $r \rightarrow \infty$ is no longer Minkowskian.
Instead,
\be
C(r) \simeq 2b \log(r) + 2b \log\log(r) + \cdots
\ee
in the limit $r\rightarrow \infty$.
However,
for small $b$,
this approximation only applies at extremely large $r$
($r$ of order ${\cal O}(e^{1/b})$ or larger).
If we restrict ourselves to a much smaller neighborhood
that is still much larger than the Schwarzschild radius,
we can still think of the Schwarzschild metric as
the approximate solution in the large $r$ limit.

Let us now study the asymptotic behavior of the solution to the Einstein equation
as we zoom into a small neighborhood of the Schwarzschild radius.
From the Einstein equations, 
we obtain 
\begin{equation}
 F = 2C \sqrt{ \frac{C+b}{4C^2 + 4 r C C' + \alpha C^{\prime\,2}}} \ .
 \label{F-sol-3}
\end{equation}
Plugging it back to the Einstein equation, 
we find
\begin{align}
 0 &= C'(r)^2 \left[\alpha  r C'(r)- 4 b \left(r^2-\alpha \right)\right] 
+4 C(r)^2 \left[\left(r^2-\alpha \right) C''(r)+2 r C'(r) - 2b\right] 
\notag\\&\qquad
+ C(r) \left[4 b \left(r^2 - \alpha\right) C''(r) - 4 b r C'(r) +6 \alpha  C'(r)^2\right] \ .
\label{c-eq-HH}
\end{align}
The perturbative expansions
\begin{align}
 C(r) &= C_0(r) + \alpha C_1(r) + \cdots \ , 
\\
 b &= \alpha b_1 + \cdots \ , 
\end{align}
give the solution for \eqref{c-eq-HH} as 
\begin{align}
 C_0(r) 
 &= 
 1 - \frac{a_0}{r} \ , 
\\
 C_1(r) 
 &= 
 - \frac{(2r-a_0)^2}{8 a_0 r^3} - \frac{c_1 + a_0 b_1}{r} 
 + \frac{2r-3a_0}{4 a_0^2 r} \left[\log r - (1-4 a_0^2 b_1) \log(r-a_0)\right] \ ,
\end{align}
where the terms inversely proportional to $r$ in $C_1(r)$ 
can be absorbed in a shift of the Schwarzschild radius $a_0$ in $C_0(r)$
by an order-$\alpha$ correction.
The next-to-leading order term $C_1(r)$ diverges except for 
\begin{equation}
 b_1 = \frac{1}{4 a_0^2} \ . 
\label{HH-pert}
\end{equation}
This is the condition on the energy flux at spatial infinities
for the Hartle-Hawking vacuum.

In addition to the perturbative approach via expansions in Newton's constant,
we shall also study the near-horizon geometry of the Hartle-Hawking vacuum
that is non-perturbative in $\alpha$ in the limit $r \rightarrow a$.
If there is a Killing horizon,
i.e.,
$C$ has a zero at $r=a$,
we assume that 
\begin{equation}
 C(r) = c_0 (r-a)^n + \cdots
\end{equation}
for some constant $n > 0$,
and then eq.\eqref{c-eq-HH} can be expanded as 
\begin{equation}
 0 = (r-a)^{2n-2}\left[4 (a^2-\alpha) b c_0^2 n + \mathcal O(r-a)\right] 
 - (r-a)^{3n-3}\left(\alpha a c_0^3 n^3 + \mathcal O(r-a)\right) \ .
\end{equation}
To satisfy this equation,
the term of order $\mathcal O((r-a)^{2n-2})$ and that of order $\mathcal O((r-a)^{3n-3})$ must cancel.
Hence 
\be
n=1 \ ,
\ee
and the equation becomes
\begin{equation}
 0 = c_0^2 (4 \alpha b - 4 a^2 b + \alpha a c_0) + \mathcal O(r-a) \ .
\end{equation}
Therefore,
$C(r)$ has a zero only if 
\begin{equation}
 b = \frac{c_0 \alpha a}{4 (a^2 - \alpha)} \ , 
\label{cond-HH}
\end{equation}
which is consistent with the perturbative result \eqref{HH-pert}. 
This is the condition for the existence of horizon.
In this case, $F$ is given by 
\begin{equation}
 F = 2 \sqrt{\frac{b}{\a}} \, (r-a) + \mathcal O((r-a)^2) \ . 
\end{equation}
As the classical Schwarzschild solution,
the near-horizon geometry for the Hartle-Hawking vacuum is given by the Rindler space.

Note that the condition \eqref{cond-HH} requires a fine-tunning of the value of $b$.
Hence it establishes a connection between the existence of horizon
and the magnitude of Hawking radiation.

Next,
consider the case when there is no horizon,
that is, 
$C(r)$ does not go to zero,
although $\rho'(r)$ diverges at some point $r=a$.
In the limit $r \rightarrow a$,
we can expand $C(r)$ as
\begin{equation}
 C(r) = c_0 + c_1 (r-a)^n + \cdots \ , 
\label{C=c0+c1}
\end{equation}
Then, the Einstein equation is expanded as 
\begin{align}
 0 &= 
 8 b c_0^2 
 + (r-a)^{n-2} \left[4 (a^2 - \alpha) (c_0 + b) c_0 c_1 n (n-1) + \mathcal O(r-a)\right] 
 + \mathcal O((r-a)^{2n-2}) 
\notag\\&\qquad
 - (r-a)^{3n-3} \left(\alpha a c_1^3 n^3 + \mathcal O(r-a)\right) \ .
\end{align}
The assumption that $\rho'(r)$ diverges at $r=a$ implies that $n < 1$, 
hence the term of order $\mathcal O((r-a)^{n-2})$ and the term of order $\mathcal O((r-a)^{3n-3})$
must cancel each other, 
so we need 
\be
n=1/2 \ .
\ee
The equation above is expanded as 
\begin{equation}
 0 = (r-a)^{-3/2} \left[\frac{1}{8} \alpha a c_1^3 - (a^2 - \alpha) (c_0 + b) c_0 c_1\right] 
 + \mathcal O\left(\frac{1}{r-a}\right) \ .
\end{equation}
It determines $c_1$ as
\begin{equation}
 c_1 = \sqrt{ \frac{8 (a^2 - \alpha)(c_0 + b)c_0}{\alpha a}} \ .
\end{equation}
The ratio $c_0/c_1$ restricts the range of validity for the approximation \eqref{C=c0+c1} to
the region \eqref{domain}.
One can then estimate $c_0$ as
\be
c_0 \leq {\cal O}\left(\frac{\a}{a^2}\right)
\ee
by matching $C(r)$ \eqref{C=c0+c1} around the point $r - a \sim {\cal O}(\a/a)$
with the Schwarzschild solution.

We use eq.\eqref{F-sol-3} to compute $F$ and find
\begin{equation}
 F = \sqrt{\frac{2 a c_0}{a^2 - \alpha}} \, \sqrt{r-a} + \mathcal O(r-a) 
\end{equation}
in the limit $r \rightarrow a$.
As we have seen in the previous section, 
this solution describes the wormhole-like geometry
in a small neighborhood of $r=a$.  

To summarize this subsection,
the horizon is possible only if $b$ is fine-tuned to the value given by eq.\eqref{cond-HH}. 
In general,
there is a wormhole solution for arbitrary non-negative $b$, 
including the case \eqref{cond-HH}. 
In the wormhole-like solution,
$\langle T^{(4)}_{uu} \rangle$ is non-zero and negative at $r=a$:
\begin{equation}
 \langle T^{(4)}_{uu}(a) \rangle = - \frac{c_0}{2 \alpha (a^2-\alpha)} \ .
\end{equation}
Its order of magnitude is ${\cal O}(1/a^4)$.
When there is a horizon, $\langle T^{(4)}_{uu} \rangle$ vanishes at the horizon.

\section{4D Scalars as Dilaton-Coupled 2D Scalars}
\label{dilaton-coupled-static}

In this section,
we consider the 2-dimensional dilaton-coupled scalar \eqref{scalar-action},
which is the dimensionally reduced 4-dimensional scalar with spherical symmetry.
Due to the coupling with dilaton,
the Weyl anomaly acquires additional terms as
\cite{Mukhanov:1994ax}
\begin{equation}
 \langle T^{(2)\,\mu}{}_\mu \rangle 
 = 
 \frac{1}{24\pi} \left[R^{(2)} - 6 (\partial\phi)^2 + 6 \nabla^2\phi\right] \ ,
\label{Weyl-dilaton}
\end{equation}
where $\mu$ is a 2-dimensional Lorentz index.

We shall consider the back reaction of the energy-momentum tensor with this anomaly,
and assume that there is no incoming or outgoing flux at spatial infinity. 
However,
the 4-dimensional conservation law \eqref{4D-cons} and the Weyl anomaly \eqref{Weyl-dilaton}
do not uniquely fix the energy-momentum tensor, 
leaving one degree of freedom unfixed.
One needs to impose an additional condition on the vacuum energy-momentum tensor,
corresponding to the choice of a quantum state.
We shall consider three possible choices:
(1) $\langle T^{(4)}_{\th\th} \rangle = \langle T^{(4)}_{\phi\phi} \rangle = 0$ (Sec.~\ref{Tthth=0}),
(2) $\langle T^{(4)}_{uu} \rangle = \langle T^{(4)}_{vv} \rangle = 0$ (Sec.~\ref{Tuu=0}),
and 
(3) the energy-momentum tensor according to Ref.\cite{Fabbri:2003fa} (Sec.~\ref{VNE}).

\subsection{Case I: $\langle T_{\th\th}^{(4)} \rangle = \langle T_{\phi\phi}^{(4)} \rangle = 0$}
\label{Tthth=0}

We first consider the vacuum state in which 
the energy-momentum tensor satisfies the 2-dimensional conservation law \eqref{2D-Conserv},
as well as the 4-dimensional one \eqref{4D-cons}. 
This implies that the angular components of the energy-momentum tensor vanish identically,
\begin{equation}
 \langle T^{(4)}_{\theta\theta} \rangle = \langle T^{(4)}_{\phi\phi} \rangle = 0 \ , 
 \label{eq:Tthth=0}
\end{equation}
as in the previous section. 

In this case,
the angular components of the Einstein equation, 
or equivalently,
the equation of motion for the dilaton $\phi$ 
is
\begin{equation}
 2\nabla^2 \phi - 2 (\partial\phi)^2 + R^{(2)} = 0 \ . 
\end{equation}
The Weyl anomaly \eqref{Weyl-dilaton} is thus simplified to
\begin{equation}
 \langle T^{(2)\,\mu}{}_\mu \rangle 
 = 
 - \frac{1}{12\pi} R^{(2)} \ , 
\end{equation}
which takes the same form as \eqref{Weyl-anomaly} 
but with an additional overall factor of $-2$. 

The energy-momentum tensor is now completely fixed by the conservation law.
It has the same forms as that of the toy model,
i.e. \eqref{theta-uu}--\eqref{theta-uv}, 
but with additional overall factors of $-2$. 
The extra factor of $-2$ can be absorbed in a redefinition of the parameter $\alpha$:
\begin{equation}
 \alpha = - \frac{\kappa N}{12\pi} \ ,
\end{equation}
which is now negative,
and then the equations in the previous section,
e.g. \eqref{F-from-C}--\eqref{rho-def},
remain formally the same.

Because of the change in sign of the parameter $\alpha$, 
we expect that the energy-momentum tensor outside the star be positive,
and the behavior of the solution near the Schwarzschild radius
can be quite different from the toy model in Sec.~\ref{ToyModel}. 
In order for the horizon or the wormhole-like geometry to appear at $r = a$,
we need
\begin{align}
 \rho'(r) &\to \infty \ , 
 \label{rhoprimediverge}
\end{align}
in the limit of $r\to a$,
which implies that
\begin{align}
 \rho''(r) &\to -\infty
\label{rho-limit}
\end{align}
in the limit.
However, eqs.\eqref{rhoprimediverge} and \eqref{rho-limit} are inconsistent with 
the Einstein equation \eqref{f-eq}.
Eq.\eqref{rhoprimediverge} implies that
eq.\eqref{region} holds 
when $r$ is sufficiently close to $a$,
so that eq.\eqref{f-eq} can be approximated by \eqref{approx-eq}.
Yet eq.\eqref{approx-eq} implies that 
$\rho''$ must be positive for $\alpha<0$ and $r^2 > \alpha$. 

The condition \eqref{rhoprimediverge} can therefore never be satisfied.
As we gradually decrease $r$,
the value of $\rho'$ increases only when $r$ is sufficiently large.
But the value of $\rho'$ starts to decrease with $r$
before it is large enough to satisfy the condition \eqref{region}.
It is therefore inconsistent to assume the existence 
of a horizon or a wormhole for the quantum state satisfying
the condition \eqref{eq:Tthth=0}.

\begin{figure}
\begin{center}
\includegraphics[scale=0.7,bb=0 0 259 170]{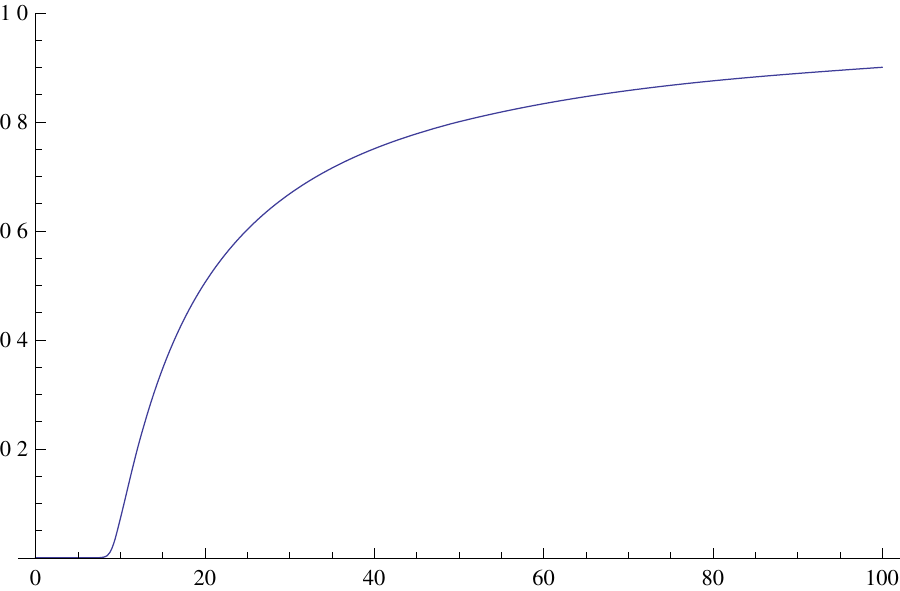}
\\
\includegraphics[scale=0.7,bb=0 0 259 170]{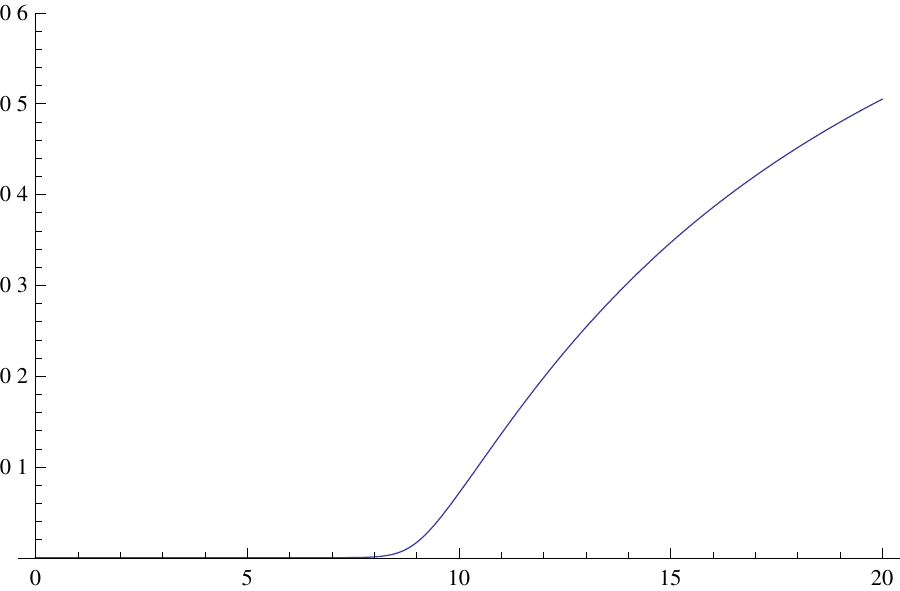}
\hspace{24pt}
\includegraphics[scale=0.7,bb=0 0 259 170]{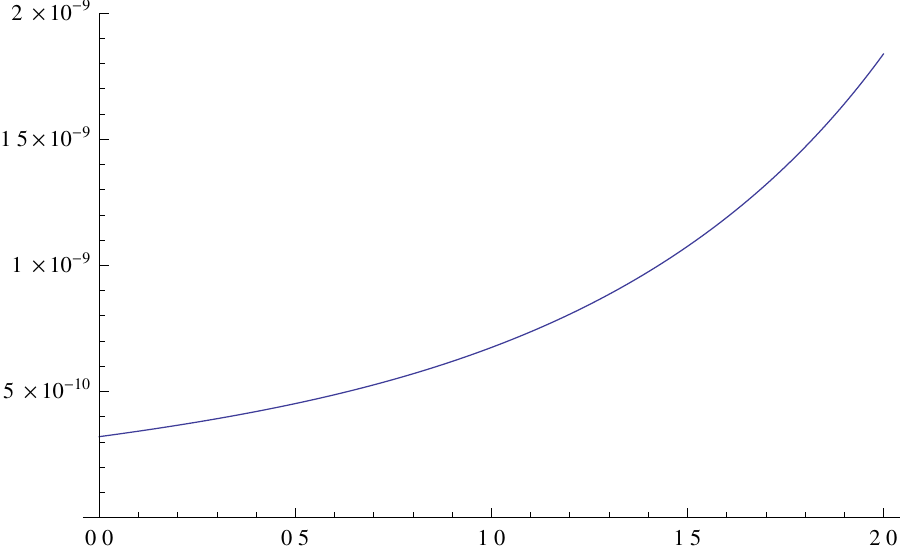}
\caption{\small Numerical results for $C(r)$. 
Top: $C(r)$ vs. $r$ from $r = 0$ to $r \gg a$.
Left: $C(r)$ vs. $r$ in a small neighborhod near $r=a$.
Right: $C(r)$ vs. $r$ near $r=0$.
$C(r)$ is always positive.}
\label{fig:C-r-2}
\end{center}
\end{figure}

\begin{figure}
\begin{center}
\includegraphics[scale=0.7,bb=0 0 259 170]{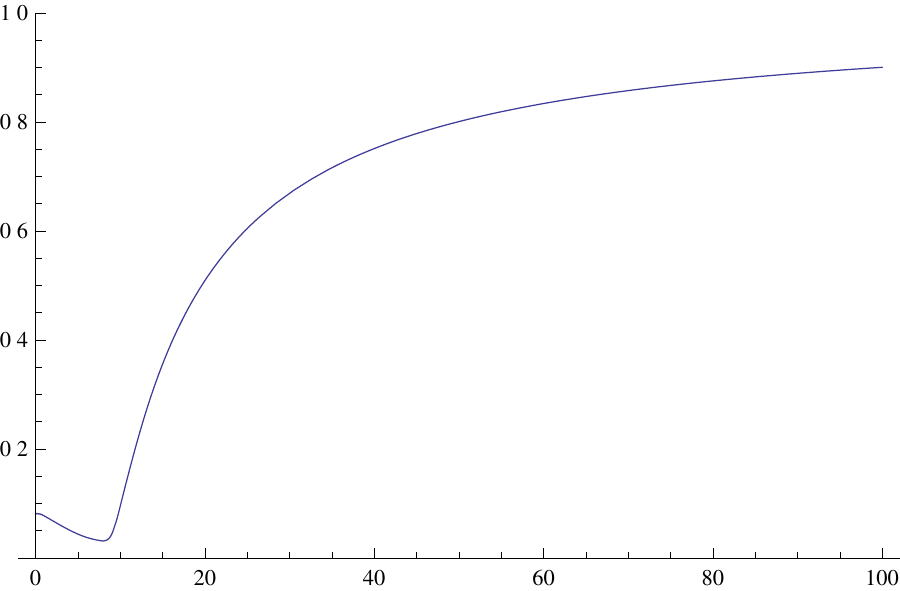}
\hspace{24pt}
\includegraphics[scale=0.7,bb=0 0 259 170]{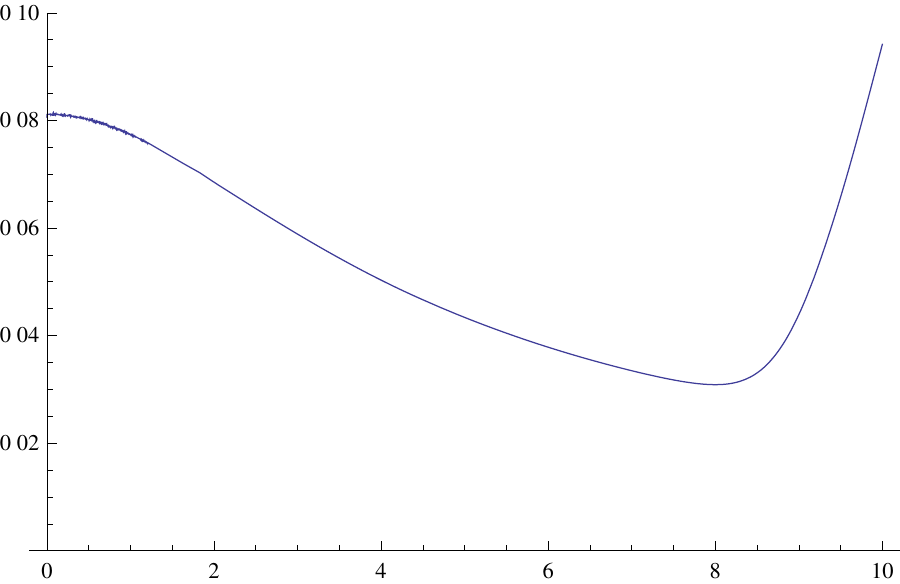}
\caption{\small Numerical results for $F(r)$. 
Left: $F(r)$ vs. $r$ from $r = 0$ to $r \gg a$.
Right: $F(r)$ vs. $r$ in a neighborhood of $r=a$.
$F(r)$ is always positive.}
\label{fig:F-r-2}
\end{center}
\end{figure}

In support of our analysis,
the numerical solutions to the Einstein equation 
are shown in Fig.\ref{fig:C-r-2} for $C(r)$
and Fig.\ref{fig:F-r-2} for $F(r)$.
As $F(r)$ is always positive,
the value of $r$ has no local minimum.
In this sense it is not like a wormhole,
but only a throat that gets narrower and narrower 
as one falls towards the center.
There is no horizon either as $C(r)$ is always positive.
Nevertheless,
$C(r)$ is extremely small for $r \sim a$ and $r < a$,
so there is a huge blue-shift for a distant observer.
Everything close to or inside the Schwarzschild radius
appears to be nearly frozen,
and it is hard to be distinguished from a real black hole
from the viewpoint of a distant observer.

\subsection{Case \II: $\langle T^{(4)}_{uu} \rangle = \langle T^{(4)}_{vv} \rangle = 0$}
\label{Tuu=0}

As another example, we impose the condition 
\begin{equation}
 \langle T^{(4)}_{uu} \rangle = \langle T^{(4)}_{vv} \rangle = 0
\end{equation}
by hand
and investigate the corresponding geometry. 
The 4-dimensional conservation law implies
\begin{equation}
 \partial_r \left(\frac{\langle T^{(2)}_{uv} \rangle}{C}\right) - 2 r \langle T^{(4)\theta}{}_\theta \rangle = 0 \ ,
\end{equation}
which determines $\langle T^{(4)\theta}{}_\theta \rangle$ in terms of $\langle T^{(2)}_{uv} \rangle$.

In this case, the equations of motion are given by 
\begin{align}
 0 &= F C' - F' C \ ,
\label{eq1dh}
\\
 0 &= r C^2 \left( C + F^2 + r FF' \right) 
 + \alpha F \left( - 6 C^2 F' + r CC' F' + r F C^{\prime\,2} - r F C C''\right)
 \ . 
\label{eq2dh}
\end{align}
We first solve these equations for $F(r)$ and obtain 
\begin{equation}
 F(r) = \frac{C(r)}{\sqrt{C(r) + r C'(r) + 6 \alpha r^{-1} C'(r) + \alpha C''(r)}} \ . 
\end{equation}
Plugging this back into \eqref{eq1dh} or \eqref{eq2dh}, 
we obtain the differential equation for $C(r)$:
\begin{equation}
 \alpha r^2 C'''(r) + (r^3 + 6 \alpha r) C''(r) + (2 r^2 + 6 \alpha) C'(r) = 0 \ . 
\end{equation}
The solution of this equation is given by 
\begin{align}
 C(r) = 1 
 - \frac{1}{r^5} 
 \left[
  \left(r^4 - 2 \alpha r^2 + 3 \alpha^2 \right) 
 \left(c_1 - c_2 \sqrt{\pi} \, \mathrm{erfc} ((2 \alpha)^{-1/2} r) \right) 
 + \sqrt{2 \alpha} \, \left(r^2 - 3 \alpha\right) e^{- \frac{r^2}{2\alpha}} 
 \right] \ , 
\label{sol-dh}
\end{align}
where erfc is the complementary error function, which is defined by 
\begin{equation}
 \mathrm{erfc} (x) = \frac{2}{\sqrt{\pi}} \int_x^{\infty} e^{-t^2} dt \ , 
\end{equation}
and $c_1$ and $c_2$ are integration constants. 
We have chosen the other integration constant such that 
$C(r) \to 1$ in the limit $r\to\infty$. 

The solution \eqref{sol-dh} has zeros for suitable choices
of the parameters $c_1$ and $c_2$. 
For example, for $c_2=0$, 
the radius of the horizon is given by 
a solution of  
\begin{equation}
 (15 \alpha^2 - 6 \alpha + 1) r^5 - c_1 r^4 + 2 \alpha c_1 r^2 - 3 \alpha^2 c_1 = 0 \ . 
\end{equation}
Since $C(r)$ behaves in the limit $r\to\infty$ as 
\begin{equation}
 C(r) \simeq 1 - \frac{c_1}{r} + \cdots 
 + \frac{16 \alpha^{5/2} c_2}{r^6} e^{- \frac{r^2}{2 \alpha}} + \cdots \ ,
\end{equation}
the constant $c_1$ is related to the mass of the black hole. 
The other constant $c_2$ specifies the quantum correction,
as it is suppressed in the limit $\alpha\to 0$, 
and hence it is not related to the classical configuration,
but a parameter for different vacua.

\subsection{Case \III}
\label{VNE}

In this subsection,
the components $\langle T_{uu}^{(2)} \rangle$ and $\langle T_{vv}^{(2)} \rangle$ 
of the energy-momentum tensor for the 2D dilaton-coupled scalar field
are calculated 
using the formula derived in Ref.\cite{Fabbri:2003fa}:
\begin{align}
 \langle T_{uu}^{(2)} \rangle 
 &=- \frac{1}{12\pi} \left( \del_u \rho \del_u \rho - \del^2_u \rho \right)
 + \frac{1}{2\pi} \left( \del_u \rho \del_u \phi + \rho (\del_u \phi)^2 \right) \ ,
\label{EE-uu-DC}
\\
 \langle T_{vv}^{(2)} \rangle 
 &=
 - \frac{1}{12\pi} \left( \del_v \rho \del_v \rho - \del^2_v \rho \right)
 + \frac{1}{2\pi} \left( \del_v \rho \del_v \phi + \rho (\del_v \phi)^2 \right) \ ,
\label{EE-vv-DC}
\end{align}
where $\rho$ is defined by \eqref{rho-def} and $\phi$ by 
\be
\phi = - \log(r/\mu) \ . 
\ee
The trace anomaly \eqref{Weyl-dilaton} is expressed in terms of $\phi$ and $\rho$ as 
\begin{equation}
 \langle T_{uv}^{(2)} \rangle = - \frac{1}{12\pi} \left( \del_u \del_v \rho
 + 3 \del_u \phi \del_v \phi - 3 \del_u \del_v \phi \right) \ .
\label{Weyl-dilaton-2}
\end{equation}
The angular components of the energy-momentum tensor is 
now non-zero and is determined through the 4-dimensional conservation law \eqref{Cons-4D-2D} 
by the rest of the energy-momentum tensor \eqref{EE-uu-DC}-\eqref{Weyl-dilaton-2}. 

The energy-momentum tensor \eqref{EE-uu-DC}-\eqref{Weyl-dilaton-2} 
can be rewritten in terms of $\rho$ and $F$ as 
\begin{align}
 \langle T^{(2)}_{uu} \rangle 
 &= 
 \frac{F(r)}{192\pi} 
 \left[
  F'(r) \rho'(r) + F(r)\left(-\rho^{\prime\,2}(r) + \rho''(r)\right)
  + \frac{6}{r^2}F(r)\left(\rho(r)-r\rho'(r)\right)
 \right] \ ,
\\
 \langle T^{(2)}_{vv} \rangle 
 &= 
 \frac{F(r)}{192\pi} 
 \left[
  F'(r) \rho'(r) + F(r)\left(-\rho^{\prime\,2}(r) + \rho''(r)\right)
  + \frac{6}{r^2}F(r)\left(\rho(r)-r\rho'(r)\right)
 \right] \ ,
\\
 \langle T^{(2)}_{uv} \rangle 
 &= 
 \frac{F(r)}{192\pi} 
 \left[
  F'(r) \rho'(r) + F(r) \rho''(r) 
  + \frac{3F'(r)}{r}
 \right] \ . 
\end{align}
By using these expressions together with 
those for the Einstein tensor
\eqref{G4uu-1}-\eqref{G4uv-1}, 
the semi-classical Einstein equation \eqref{Einstein-2D} gives 
the following differential equations:
\begin{align}
 0 &= 
 - r^2 F'(r) \left(2 \alpha \rho'(r) + r\right) 
\notag\\&\qquad 
 - 2 F(r) 
 \left[
  \alpha r^2 \rho''(r) - \alpha r^2 \rho^{\prime\,2}(r) - r (r^2 - 6\alpha) \rho'(r) 
  + 6 \alpha\rho(r)  
 \right] \ , 
\label{eq1d}
\\
 0 &= 
 e^{2\rho(r)} 
 - F(r)F'(r) \left(2 \alpha r\rho'(r) + r^2 + 6\alpha\right) 
 - F^2(r) \left(2 \alpha r \rho''(r) + r\right) \ . 
\label{eq2d}
\end{align}
From these differential equations, 
we can easily solve $F(r)$ as 
\begin{equation}
 F(r) = e^{\rho(r)} r^{3/2} \sqrt{ \frac{2 \alpha\rho'(r)+r}{D(r)}} \ , 
\label{F-dilaton}
\end{equation}
where the function $D(r)$ is 
\begin{align}
 D(r) &= 
 r^4 - 12 \alpha^2 r^2 \rho''(r) 
 - 12 \alpha \rho(r) \left(2 \alpha r\rho'(r) + r^2 + 6\alpha\right) 
\notag\\&\qquad
 + 2 \rho'(r) 
 \left[
 \alpha r^2 \rho'(r) \left(2 \alpha r\rho'(r) + 3 (r^2 + 6\alpha)\right) 
 + r (r^2 + 4\alpha)(r^2 + 9\alpha)\right] \ . 
\end{align}
Plugging \eqref{F-dilaton} back into \eqref{eq1d} or \eqref{eq2d}, 
we obtain the differential equation for $\rho(r)$:
\begin{align}
 0 &= 
 - 24 \alpha^2 r^2 \rho (r) \rho ''(r) \left(15 \alpha +r^2+2 \alpha  r
   \rho '(r)\right) - 144 \alpha ^2 r^2 \rho (r)
\notag\\&\quad
 + 12 \alpha  r \rho (r) \rho '(r) \left[4 \alpha  r \rho '(r)
   \left(14 \alpha + 2 r^2 +\alpha  r \rho '(r) \right) + 3 r^4+ 40 \alpha r^2 + 126 \alpha ^2\right]
\notag\\&\quad
 + 2 \alpha  r^3 \left(186 \alpha ^2+3 r^4+56 \alpha  r^2\right) \rho
   '(r)^3
 + 4 \alpha ^2 r^4 \left(12 \alpha +r^2\right) \rho '(r)^4 
\notag\\&\quad
 + 2 r^2 \rho '(r)^2 \left(324
   \alpha ^3+r^6+27 \alpha  r^4-18 \alpha ^3 r^2 \rho ''(r)+162 \alpha ^2
   r^2\right)
\notag\\&\quad
 -6 \alpha ^2 r^5 \rho ^{(3)}(r) 
 + r^4 \rho ''(r) \left(48 \alpha ^2+r^4+10 \alpha r^2+36 \alpha ^3 \rho ''(r)\right) 
\notag\\&\quad
 +2 r^3 \rho '(r) \left(-72 \alpha ^2+r^4-3
   \alpha  r^2+\alpha  \left(-138 \alpha ^2+r^4-14 \alpha  r^2\right) \rho ''(r)-6
   \alpha ^3 r \rho ^{(3)}(r)\right) \ . 
\label{eq3d}
\end{align}

If there is a Killing horizon at $r=a$, 
we must have $\rho\to -\infty$ as $r\to a$. 
Then $\rho$ would behave around $r=a$ as either
\begin{equation}
 \rho(r) = \rho_0 \log(r-a) + \cdots \ , \label{rho-log}
\end{equation}
or 
\begin{equation}
 \rho(r) = \frac{1}{2}\log c_0 + \rho_0 (r-a)^n + \cdots
 \label{rho-power}
\end{equation}
with $n<0$. 

Assuming eq.\eqref{rho-log}, which includes the case of the Schwarzschild solution,
the Einstein equation \eqref{eq3d} can be expanded as 
\begin{equation}
 0 = 4 \alpha^2 a^4 \rho_0^2 
 \left(a^2 \rho_0^2 + 12 \alpha \rho_0^2 + 9 \alpha \rho_0 + 3 \alpha \right) \frac{1}{(r-a)^4} 
 + \mathcal O\left(\frac{1}{(r-a)^3}\right) \ ,
\end{equation}
and we can solve $\rho_0$ as
\begin{equation}
 \rho_0 = \frac{1}{2 a^2 + 24 \alpha} 
 \left(- 9 \alpha \pm \sqrt{- 12 \alpha a^2 - 63 \alpha^2}\right) \ , 
\end{equation}
which is never real since $a^2 \gg \a$.
Therefore,
$\rho$ can never behave as \eqref{rho-log} near $r=a$. 

For the other option \eqref{rho-power}, 
the Einstein equation \eqref{eq3d} is expanded as 
\begin{align}
 0 &= 
 36 \alpha^2 \log c_0 \left[2a^2 + (a^2 + 6 \alpha)\log c_0 \right] + \mathcal O(r-a) 
\notag\\&\quad
 + (r-a)^{n-3} \left[- 6 \alpha^2 a^5 n (n-1) (n-2) \rho_0 + \mathcal O(r-a)\right] 
\notag\\&\quad
 + (r-a)^{2n-4} \left[12 \alpha^3 a^4 n^2 (n-1) (2n-1) \rho_0^2 + \mathcal O(r-a)\right] 
\notag\\&\quad
 - (r-a)^{3n-4} \left[36 \alpha^3 a^4 n^3 (n-1) \rho_0^3 + \mathcal O(r-a)\right] 
\notag\\&\quad
 + (r-a)^{4n-4} \left[4 \alpha^2 a^4 (a^2 + 12 \alpha) n^4 \rho_0^4 + \mathcal O(r-a)\right]
 + \mathcal O(1) \ .
\end{align}
In order for the leading order terms to cancel,
we need
\begin{equation}
 n = \frac{1}{2} \ . 
\end{equation}
Then, $C(r)$ behaves near $r=a$ as 
\begin{equation}
 C(r) \simeq c_0 e^{2\rho_0 \sqrt{r-a}} \ . \label{C-dilaton}
\end{equation}
The coefficient $\rho_0$ can be fixed from the leading order term of 
the expansion of \eqref{eq3d} around $r=a$,
\begin{equation}
 0 = 
 \frac{9}{4} \alpha^2 a^4 \rho_0 (\alpha \rho_0^2 - a) (r-a)^{-5/2} + \mathcal O((r-a)^{-2}) \ , 
\end{equation}
to be
\begin{equation}
 \rho_0 = \sqrt{ \frac{a}{\alpha}} \ . 
\end{equation}
Using \eqref{F-dilaton} with \eqref{C-dilaton}, 
we find
\begin{equation}
 F(r) \simeq \sqrt{ \frac{2 c_0 a (r-a)}{\alpha (a\rho_0^2 + 6)}} \label{F-approx-WH-D}
\end{equation}
in the limit $r \to a$.
Since $C(r)$ is non-zero and $F(r)$ behaves as $\mathcal O(\sqrt{r-a})$ near $r=a$, 
the metric in the limit $r \to a$ is approximately given by that of the wormhole as 
in the case of Sec.~\ref{ToyModel}. 

The back reaction of vacuum energy due to dilaton-coupled 2-dimensional scalar 
has been studied previously in Ref.~\cite{Fabbri:2005nt},
which announced the absence of horizon and the existence of a ``turning point''
(i.e. $F(a) = 0$) using numerical analysis,
in agreement with our results of analytic arguments. 
They also claimed that there is a divergence of $F$ 
beyond the turning point in their numerical analysis.
Such a singularity exists only if the surface of the star is
sufficiently far away from the point $r = a$ 
so that the vacuum solution still applies to the neighborhood of the singularity.
As our focus is on the local geometry that replaces the near-horizon region,
a singularity further down the throat is not of our concern.
(See the discussion at the end of Sec. \ref{Non-Perturbative-Analysis}.)

Incidentally,
let us prove analytically that there is no singularity 
which is associated to the pole of $F(r)$. 
As we have discussed above, $C(r)$ does not have 
divergence or zero at finite and non-zero $r$. 
If there is a curvature singularity but $C(r)$ is regular there, 
$F(r)$ must diverge at the singularity. 
It was also proposed in Ref.~\cite{Fabbri:2005nt} by using numerical analyses 
that the singularity occurs at a point $r = r_M$
where $\rho$ is finite and $F$ diverges as
$
F(r) \propto (r_M - r)^{-1/2}
$
in the limit $r \rightarrow r_M$, 
(See eq.(103) in Ref.\cite{Fabbri:2005nt}.)
for the semi-classical Einstein equations (eqs.(30)-(32) in Ref.\cite{Fabbri:2005nt}) 
which is identical to \eqref{eq1d}-\eqref{eq2d} in this paper.%
\footnote{%
Eqs.(30)-(31) in Ref.\cite{Fabbri:2005nt} is expressed in terms of 
$\rho(r_*)$ and $\phi(r_*)$ while \eqref{eq1d}-\eqref{eq2d} in this paper are 
written by using $\rho(r)$ and $F(r)$. 
They are related to each others by $F=\frac{dr}{dr_*}$ and $\phi = -\log(r/\mu)$. 
Eq.(32) in Ref.\cite{Fabbri:2005nt} is obtained from the consistency condition 
with the Bianchi identity. 
} 
However, the singularity of this sort 
is incompatible with the semi-classical Einstein equations \eqref{eq1d}-\eqref{eq2d}, 
as we now prove below. 
First of all, according to \eqref{F-approx-WH-D},
$F(r)$ must be finite if $\rho'$ diverges.
This implies that 
$\rho'(r_M)$ must be finite if $F(r)$ diverges as
\begin{equation}
F(r) \propto (r_M - r)^n \ 
\label{F-r-n}
\end{equation}
for some negative $n$
($n = - 1/2$ in Ref.\cite{Fabbri:2005nt}).
The leading order terms in the Einstein equations \eqref{eq1d} and \eqref{eq2d} are then
\begin{align}
 0 &= - n (2 \alpha \rho'(r_M) + r_M) r_M^2 (r_M - r)^{n-1} \ , 
\\
 0 &= - n (2 \alpha r_M \rho'(r_M) + r_M^2 + 6 \alpha) (r_M - r)^{2n-1} \ .
\end{align}
Hence we see that the two Einstein equations are inconsistent 
with their ansatz of the singularity.
Therefore, the singularity can exist only in the limit $r\to\infty$, 
although it can be in a finite affine distance from finite $r$. 

\section{Energy-Momentum Tensor and Near-Horizon Geometry}
\label{General}

In Secs.~\ref{ToyModel}~and~\ref{dilaton-coupled-static},
we considered different models of the vacuum energy-momentum tensor,
which is always found to be regular at the horizon
(in a local orthonormal frame)
when the back reaction is taken into account.
Our opinion is that a reasonable model for the vacuum energy-momentum tensor
should prevent divergence in local orthonormal frames by itself 
at least at the macroscopic scale.
We also found that sometimes the existence of horizon demands fine-tunning,
and it can be easily deformed into a wormhole-like geometry without horizon
by a small modification of the energy-momentum tensor 
within a tiny range of space.
Our observation is that
horizons are extremely sensitive to tiny changes
in the energy-momentum tensor at the horizon.
In this section,
we zoom into the tiny space around the horizon
(or the wormhole-like space)
and explore the connection between its geometry
and the energy-momentum tensor,
without specifying any detail about the physical laws 
behind the vacuum energy-momentum tensor.

We consider the (semi-classical) Einstein equations 
for 4-dimensional static, spherically symmetric geometries 
with an arbitrary energy-momentum tensor. 
According to eqs.\eqref{G4uu-1}-\eqref{G4thth-1}, 
the Einstein 
equations
are
\begin{align}
G_{uu} &= \frac{1}{2 C(r) r} \left[ F^2 C'(r) - \frac{1}{2} C(r)(F^2)'(r) \right] 
= \kappa T_{uu} \ ,
\label{Ein-A-uu}
\\
G_{vv} &= \frac{1}{2 C(r) r} \left[ F^2 C'(r) - \frac{1}{2}C(r)(F^2)'(r) \right] 
= \kappa T_{vv} \ ,
\label{Ein-A-vv}
\\
G_{uv} &= \frac{1}{2 r^2} \left[ C(r) - F^2 - \frac{r}{2} (F^2)'(r) \right] 
= \kappa T_{uv} \ ,
\label{Ein-A-uv}
\\
G_{\theta\theta} &= 
- \frac{r^2}{2 C^3} \left[ F^2 C^{\prime\,2} - \frac{1}{2} (F^2)' C C' - F^2 C C'' \right]
+ \frac{r}{2C} (F^2)' 
= \kappa T_{\th\th} \ .
\label{Ein-A-thth}
\end{align}
Note that $F(r)$ appears only in the form of $F^2(r)$.
In this section, 
we shall omit the superscript $(4)$
while all quantities are defined in the 4-dimensional theory.
We will denote $\langle T^{(4)}_{\mu\nu} \rangle$ simply as $T_{\mu\nu}$.

For static and spherically symmetric configurations, 
the energy-momentum tensor $T_{\mu\nu}$ are functions 
which depend only on $r$. 
They allow us to solve the function $F$ as
\be
F^2(r) = \frac{2\kappa r^2(T_{uu}(r) - T_{uv}(r)) + C(r)}{2r\rho'(r) + 1} \ ,
\label{sol-F}
\ee
where $\rho(r)$ is defined by
\be
C(r) = e^{2\rho(r)} \ .
\ee
Incidentally,
as results of the Einstein equations and spherical symmetry, 
we have
\bea
G_{\th\th} &=& - r^2 R^{u}{}_{u} \ ,
\label{Ruv-T}
\\
R_{\th\th} &=& - r^2 G^{u}{}_{u} \ .
\label{Tuv-R}
\eea

The Einstein equations \eqref{Ein-A-uu} -- \eqref{Ein-A-thth},
together with the regularity of the energy-momentum tensor,
will be our basis to establish the connection between
the energy-momentum tensor and the existence of horizon.

\subsection{Conditions for Horizon}
\label{Condition-Horizon}

For static configurations
with spherical symmetry,
the event horizon and the apparent horizon coincide with the Killing horizon. 
In this subsection, we consider the metric \eqref{MetricA} with a Killing horizon at $r = a$,
so
\be
C(a) = 0 \ ,
\ee
which implies that
\be
\rho \rightarrow - \infty \ ,
\qquad
\rho' \rightarrow \infty
\ee
as $r \rightarrow a$.
Assuming that $T_{uu}$ and $T_{uv}$ are finite, 
eq.\eqref{sol-F} implies that $F(r) = 0$ at the Killing horizon. 

For solutions of the Einstein equation, 
the regularity of the geometry implies 
the regularity of the energy-momentum tensor. 
As $g^{uv} R_{uv}$ and $R_{\theta\theta}$ 
should both be regular for a regular space-time with spherical symmetry,
eqs.\eqref{Ruv-T} and \eqref{Tuv-R} say that
$g^{uv} T_{uv}$ and $T_{\theta\theta}$ should both be finite. 
Therefore, $T_{uv}$ must vanish at $r=a$ and 
it is convenient to express it in terms of $T^u{}_u = - 2 C^{-1} T_{uv}$,
which should be regular but can be non-zero at $r=a$. 

$F(r)$ (\ref{sol-F}) can thus be rewritten as
\begin{equation}
 F^2(r) = \frac{2\kappa r^2 T_{uu}(r) + C(r)(1 + \kappa r^2 T^u{}_u(r))}{2r\rho'(r) + 1} \ ,
\label{sol-F-reg}
\end{equation}
where $T^u{}_u$ is regular at $r = a$.
Since $C(a)=0$,
we assume that $C$ can be expanded as
\begin{equation}
 C(r) = c_0 (r-a)^n 
 + \cdots \ , 
\label{C-expand}
\end{equation}
in the limit $r \to a$
with $n>0$. 
Plugging \eqref{sol-F-reg} back to \eqref{Ein-A-uu} or \eqref{Ein-A-uv} 
and expand around $r=a$ by using \eqref{C-expand}, 
we obtain 
\begin{equation}
 0 = (r-a)^{2n-2}\left[- 2 \kappa a^2 c_0^2 n T_{uu}(a) + \mathcal O(r-a)\right] 
 + \mathcal O((r-a)^{3n-2}) \ . 
\end{equation}
Therefore,
the Einstein equation at the leading order implies that
$T_{uu}$ (and $T_{vv}$) must vanish at the Killing horizon $r=a$.

The condition that $T_{uu}$ and $T_{vv}$ must vanish at the horizon
can be understood as follows.
Physically,
the regularity of the energy-momentum tensor should be checked
in a local orthonormal frame.
The finiteness of $T_{uu}$ or $T_{vv}$ is not sufficient to ensure the regularity
as the coordinates $(u, v)$ are singular at the horizon
in the sense that $C(a) = 0$ \cite{Christensen:1977jc}. 

Let us now examine the regularity condition for
the energy-momentum tensor at the horizon.
At the future horizon ($du=0$), 
we should find another coordinate $\tilde u$
such that the metric is regular in the coordinate system $(\tilde{u}, v)$.
That is,
in terms of the coordinates $(\tilde{u}, v)$,
the metric becomes
\begin{equation}
 ds^2 = - \widetilde C d\tilde u \,dv + r^2 d \Omega^2 \ , 
\end{equation}
where 
\begin{equation}
 \widetilde C \equiv C \frac{du}{d\tilde u} \ , 
\end{equation}
and we need $\widetilde C$ to be finite and non-zero at $r=a$
in order for $(\tilde{u}, v)$ to be a regular local coordinate system at the horizon.
Then, we have 
\begin{equation}
 \frac{du}{d\tilde u} \propto C^{-1} \to \infty 
\end{equation}
as $r\to a$,
and therefore
\begin{align}
T_{\tilde{u}\tilde{u}} 
 &= 
 \left( \frac{du}{d\tilde{u}} \right)^2 T_{uu} \ , 
&
 T_{\tilde{u}{v}} 
 &= 
 \frac{du}{d\tilde{u}} T_{uv}
\end{align}
would both diverge at $r = a$
unless
\be
T_{uu}(a) = T_{uv}(a) = 0 \ .
\ee
Since $T_{vv}=T_{uu}$ for static configurations,
we also have $T_{vv} = 0$ at the horizon.
To be more precise, 
$T_{uu}$, $T_{vv}$ and $T_{uv}$ must behave as 
\begin{align}
 T_{uu} 
 &= \mathcal O(C^2) \ , 
& 
 T_{vv} 
 &= \mathcal O(C^2) \ , 
&
 T_{uv} 
 &= \mathcal O(C)
\label{limit-T-a}
\end{align}
as $r\to a$. 

For static geometries, 
a coordinate system which covers only the intersection of 
the future and past horizons are sometimes used. 
In this case, 
we must transform both coordinates to new coordinates $(\tilde{u}, \tilde{v})$
in order for the metric to be regular,
\begin{equation}
 ds^2 = - \widetilde C d\tilde u\,d\tilde v + r^2 d \Omega^2 \ , 
\end{equation}
where 
\begin{equation}
 \widetilde C \equiv C \frac{du}{d\tilde u} \frac{dv}{d\tilde v} \ . 
\end{equation}
In order for $\widetilde C$ to be finite and non-zero at $r=a$, 
we need
\begin{equation}
 \frac{du}{d\tilde u} \frac{dv}{d\tilde v} \propto C^{-1} \to \infty \ . 
\end{equation}
If we take $\tilde u$ and $\tilde v$ such that they are simply exchanged
(up to sign)
under the time reversal transformation, 
The energy-momentum tensor must behaves as 
\begin{align}
 T_{uu} 
 &= \mathcal O(C) \ , 
& 
 T_{vv} 
 &= \mathcal O(C) \ , 
& 
 T_{uv} 
 &= \mathcal O(C) 
\end{align}
in $r\to a$. 

This simple mathematical result can have surprising implications 
because it says that it is possible for an arbitrarily small modification 
to the energy-momentum tensor at the horizon to kill the horizon.
Conceptually, this explains why the horizon of the Schwarzschild solution disappears
when we turn on the quantum correction to the vacuum energy-momentum tensor
as we have shown in Secs.~\ref{Non-Perturbative-Analysis},~\ref{Tthth=0}~and~\ref{VNE}.
It also explains why one needs to fine-tune the additional energy flux 
in order to admit the existence of a horizon in Sec.~\ref{Hartle-Hawking-Vacuum}.


\subsection{Asymptotic Solutions in Near-Horizon Region}

In this subsection,
we shall examine more closely the relation between 
the energy-momentum tensor at the horizon
and the near-horizon geometry for
a series of near-horizon solutions.

For a generic quantum theory,
the vacuum energy-momentum tensor is typically a polynomial of
finite derivatives of the metric.
Then, 
as we have shown in the examples in Secs.\ref{ToyModel} and \ref{dilaton-coupled-static},
the Einstein equation in the limit $r \rightarrow a$ leads to
a differential equation involving only the leading order terms:
\be
(C^{(n_1)})^{m_1} + a(C^{(n_2)})^{m_2}(C^{(n_2)})^{m_3}(\cdots) \simeq 0 \ ,
\ee
where $(n_1), (n_2), (n_3)$ are the order of derivatives with respect to $r$.
If this equation admits an asymptotic solution as \eqref{C-expand},%
\footnote{
We will not consider all possible solutions.
For instance,
the solutions with $C(r) \propto \exp(-c(r-a)^{-\b})$
in the limit $r \to a$
also have horizons ($c, \b > 0$),
but will not be included in the discussions below.
}
$n$ must satisfy an algebraic equation of the form
\be
m_1(n-n_1) = m_2(n-n_2) + m_3(n-n_3) + \cdots \ .
\ee
which is always solved by a rational number
\be
n = \frac{K}{M} \ ,
\qquad
(K, M \in \mathbb{Z}).
\ee

The subleading terms in $C(r)$ \eqref{C-expand} in the limit $r \to a$
should be determined by the subleading terms in the Einstein equations.
To be sure that the leading-order solution is part of a consistent solution,
one needs a consistent expansion scheme for which
higher and higher order terms in $C(r)$ 
can be solved order by order from the Einstein equations.
In view of the Einstein equations \eqref{Ein-A-uu}--\eqref{Ein-A-thth},
it is clear that a consistent ansatz for the expansion of $C(r)$ is
\be
C(r) = (r-a)^{K/M} \left[ c_0 + c_1 (r-a)^{1/M} + c_2 (r-a)^{2/M} + \cdots \right]
\label{C-general-expansion}
\ee
for some integers $K \geq 0$ and $M \geq 1$.
Eq.\eqref{sol-F}) then implies that
\be
F^2(r) = (r-a)^{K'/M + 1} \left[ f_0^2 + \tilde f_1 (r-a)^{1/M} + \tilde f_2 (r-a)^{2/M} + \cdots \right]
\label{F2-general-expansion}
\ee
for a certain integer $K' \geq 0$.

In the limit $r \to a$,
the metric for $C(r)$ \eqref{C-general-expansion} and $F^2(r)$ \eqref{F2-general-expansion} is
\bea
ds^2 &\simeq& - c_0 (r-a)^{K/M} dt^2 + \frac{c_0}{f_0^2 (r-a)^{(M+K'-K)/M}} dr^2 + a^2 d\Omega^2
\nn \\
&\simeq& - c_0 x^2 dt^2 + \frac{4M^2 c_0}{K^2 f_0^2} \frac{dx^2}{x^{2(K'-M)/K}} + a^2 d\Omega^2 \ ,
\label{metric-general}
\eea
where $r = a + x^{2M/K}$.

Assuming that there is no other length scale except $a$ and $\alpha$,
the expansions \eqref{C-general-expansion} and \eqref{F2-general-expansion}
are expected to be valid when
\be
0 \leq r - a \ll \frac{\alpha}{a} \ .
\ee
A rough estimate of the values of $c_0$ and $f_0$ can be made 
by matching $C(r)$ and $F^2(r)$ at the leading order
with the Schwarzschild solution for $r - a \sim {\cal O}(\alpha/a)$,
if the solution is well approximated by the Schwarzschild metric at large $r$.
We find
\be
c_0 \sim {\cal O}\left(\frac{\a^{1-K/M}}{a^{2-K/M}}\right) \ ,
\qquad
f_0^2 \sim {\cal O}\left(\frac{\a^{1-K'/M}}{a^{3-K'/M}}\right) \ .
\label{c0-f0-order}
\ee

We now study the condition on the energy-momentum tensor 
in order for the horizon to exist.
The energy-momentum tensor is determined by 
$C(r)$ \eqref{C-general-expansion} and $F^2(r)$ \eqref{F2-general-expansion}
through the Einstein equations
as an expansion in powers of $(r-a)^{1/M}$:
\bea
\kappa T^{v}{}_{u}(r) &=& G^{v}_{u}
= (r-a)^{(-K+K')/M} \frac{(-2K + K' + M) f_0^2}{2M a c_0} + \cdots \ ,
\label{T4vu-general}
\\
\kappa T^{u}{}_{u}(r) &=& G^{u}_{u}
= - \frac{1}{a^2} + (r-a)^{(-K+K')/M} \frac{(K' + M) f_0^2}{2M a c_0} + \cdots \ ,
\label{T4uu-general}
\\
\kappa T_{\th\th}(r) &=& G_{\th\th}
= - (r-a)^{(-M-K+K')/M}\frac{K(M-K') a^2 f_0^2}{4M^2 c_0} + \cdots \ .
\label{T4thth-general}
\eea
Constraints should be imposed on the coefficients of the singular terms
as $T^{v}{}_{u}(r)$, $T^{u}{}_{u}(r)$ and $T_{\th\th}(r)$ should all be regular at the horizon $r = a$,
as we have argued above.

Depending on the values of $K, K'$ and $M$,
a solution can be classified into one of the following categories:

\begin{enumerate}
\item
If $K > K'$,
in order for $T^{v}{}_{u}(a)$ and $T^{u}{}_{u}(a)$ to be finite,
we need $K = 0$,
which implies that there is no horizon.
This case will be considered in the next subsection.

\item
If $K = K'$,
in order for $T_{\th\th}(a)$ to be finite,
we need $M = K'$
(and there are more constraints on the coefficients 
in the expansions of $C(r)$ \eqref{C-general-expansion}
and $F^2(r)$ \eqref{F2-general-expansion}
if $M > 1$).
In such cases,
\bea
\kappa T^{v}{}_{u}(a) &=& G^{v}{}_{u}
= 0 \ ,
\\
\kappa T^{u}{}_{u}(a) &=& G^{u}{}_{u}
= - \frac{1}{a^2} + \frac{f_0^2}{a c_0} > - \frac{1}{a^2} \ ,
\\
\kappa T_{\th\th}(a) &=& G_{\th\th}
= \mbox{depends on $M$} \ ,
\\
ds^2 &\simeq& - c_0 (r-a) dt^2 + \frac{c_0}{f_0^2 (r-a)} dr^2 + a^2 d\Omega^2
\nn \\
&\simeq& - c_0 x^2 dt^2 + \frac{4M^2 c_0}{K^2 f_0^2} dx^2 + a^2 d\Omega^2 \ ,
\eea
where
$\frac{f_0^2}{ac_0} \sim {\cal O}\left(\frac{1}{a^2}\right)$
and $r = a + x^{2}$.
The near-horizon geometry is the Rindler space.
This case includes the classical Schwarzschild solution
and the Hartle-Hawking vacuum considered in Sec.~\ref{Hartle-Hawking-Vacuum}.
Note that $f_0^2/c_0$ is of order ${\cal O}(1/a)$,
hence $G^{u}{}_{u}(a)$ is of order ${\cal O}(1/a^2)$.

\item
If $K < K'$ and $M > (K'-K)$,
in order for $T_{\th\th}(a)$ to be finite,
we need $M = K'$
(and there are more constraints on the coefficients 
in the expansions of $C(r)$ \eqref{C-general-expansion}
and $F^2(r)$ \eqref{F2-general-expansion}
if $M > K'-K+1$).
In such cases,
\bea
\kappa T^{v}{}_{u}(a) &=& G^{v}{}_{u}
= 0 \ ,
\\
\kappa T^{u}{}_{u}(a) &=& G^{u}{}_{u}
= - \frac{1}{a^2} \ ,
\\
\kappa T_{\th\th}(a) &=& G_{\th\th}
= \mbox{depends on $M$} \ ,
\\
ds^2 &\simeq& - c_0 (r-a)^{K/M} dt^2 + \frac{c_0}{f_0^2 (r-a)^{(M+K'-K)/M}} dr^2 + a^2 d\Omega^2
\nn \\
&\simeq& - c_0 x^2 dt^2 + \frac{4M^2 c_0}{K^2 f_0^2} dx^2 + a^2 d\Omega^2 \ ,
\eea
where $r = a + x^{2M/K}$.
Again we have the Rindler space.

\item
If $K < K'$ and $M = (K'-K)$,
\bea
\kappa T^{v}{}_{u}(a) &=& G^{v}{}_{u}
= 0 \ ,
\\
\kappa T^{u}{}_{u}(a) &=& G^{u}{}_{u}
= - \frac{1}{a^2} \ ,
\\
\kappa T_{\th\th}(a) &=& G_{\th\th}
= \frac{K^2 a^2 f_0^2}{4M^2 c_0} > 0 \ ,
\\
ds^2 &\simeq& - c_0 (r-a)^{K/M} dt^2 + \frac{c_0}{f_0^2 (r-a)^{(M+K'-K)/M}} dr^2 + a^2 d\Omega^2
\nn \\
&\simeq& - c_0 x^2 dt^2 + \frac{4M^2 c_0}{K^2 f_0^2} \frac{dx^2}{x^{2}} + a^2 d\Omega^2 \ ,
\eea
where $r = a + x^{2M/K}$.
This metric describes $AdS_2\times S^2$,
which is the near horizon geometry of the extremal Reissner-Nordstr\"{o}m black hole.
The order of magnitude of $G^{\th}{}_{\th}(a)$ is ${\cal O}(1/a^2)$.%
\footnote{
We can no longer use the estimate \eqref{c0-f0-order},
which assumes that the metric is Schwarzschild at larger $r$.
The estimate here is done by assuming the extremal RN black hole metric at large $r$.
}

\item
If $K < K'$ and $M < (K' - K)$,
\bea
\kappa T^{v}{}_{u}(a) &=& G^{v}{}_{u}
= 0 \ ,
\\
\kappa T^{u}{}_{u}(a) &=& G^{u}{}_{u}
= - \frac{1}{a^2} \ ,
\\
\kappa T_{\th\th}(a) &=& G_{\th\th}
= 0 \ ,
\\
ds^2 &\simeq& - c_0 (r-a)^{K/M} dt^2 + \frac{c_0}{f_0^2 (r-a)^{(M+K'-K)/M}} dr^2 + a^2 d\Omega^2
\nn \\
&\simeq& - c_0 x^2 dt^2 + \frac{4M^2 c_0}{K^2 f_0^2} \frac{dx^2}{x^{2(K'-M)/K}} + a^2 d\Omega^2 \ ,
\eea
where $r = a + x^{2M/K}$.
As in the previous cases,
it takes an infinite amount of time (change in $t$)
to reach the horizon at $r = a$ from the viewpoint of a distant observer.
\end{enumerate}

For all of the near-horizon geometries,
we find
\be
\kappa T^{v}{}_{u}(a) = G^{v}{}_{u}(a) = 0 \ ,
\quad \mbox{and} \quad
\kappa T^{u}{}_{u}(a) = G^{u}{}_{u}(a) \geq - \frac{1}{\kappa a^2} \ .
\ee
They imply that there is no Killing horizon if $T_{uu}$ or $T_{uv}$ is non-zero.
While the first condition was derived in Sec.\ref{Condition-Horizon},
the second condition arises only after a detailed analysis.

We should emphasize here that the solutions above 
may or may not be extended beyond the point $r = a$
without singularity.
For our purpose to investigate common features of solutions with horizon,
we aim at including as many possibilities as possible.

\subsection{Absence of Horizon}

In this subsection,
we consider the connection between wormhole-like geometry without horizon
and the energy-momentum tensor.
The stereotype of a traversable wormhole is a smooth structure that
connects two asymptotically flat spaces,
allowing objects to travel from one side to the other.
Its cross sections are 2-spheres,
whose area is typically minimized in the middle of the connection (``throat'').
In particular,
a 3-dimensional spherically symmetric space 
can be viewed as a foliation of concentric 2-spheres.
The surface area of the 2-sphere depends on the distance
between the center and the points on the 2-sphere,
although the latter is not necessarily a monotonically increasing function of the former.

For the metric \eqref{MetricA},
the area of the 2-sphere is $4\pi r^2$.
By a ``wormhole-like geometry'',
we mean the existence of a local minimum in the value of $r$,
identified as the narrowest point of the throat of the wormhole.
It is not a genuine wormhole because 
only one side of the throat is an open space,
while the other side is expected to be closed,
filled with matter of positive energy around the origin.

Another type of peculiar geometry that will also be considered below
is the limit of the wormhole-like geometry
in which the throat is infinitely long.

Assuming that there is a wormhole-like geometry
with the local minimal value of the function $r$ equal to $a$,
we expect that $dr/dr_* = 0$
\footnote{
It is however not true that
the condition $dr/dr_* = 0$ always implies a local minimum of $r$.
}
and thus $F(r) = 0$ at $r = a$.
The condition $F(r) = 0$ will also be satisfied at $r = a$
in the limit of an infinitely long throat.
In the limit $r \rightarrow a$,
the wormhole-like metric is of the form:
\be
ds^2 \simeq - C(a) (dt^2 - dr_{\ast}^2) + a^2 d\Omega^2 \ ,
\ee
describing a neighborhood of $r = a$ with the topology $R^2 \times S^2$.
This resembles a traversable wormhole,
although it terminates at the surface of a star rather than
leading to an open space.
It is relevant only when the radius of the star is smaller than
the Schwarzschild radius.

If $F(a)=0$ but $C(a) \neq 0$,
there is no horizon at $r = a$.
According to \eqref{sol-F}, 
in order for $F(r)$ to vanish,
either $\rho'(r)$ diverges at $r = a$,
or the energy-momentum tensor satisfies the condition
\be
T_{uu}(a) - T_{uv}(a) = - \frac{C(a)}{2\kappa a^2} \ .
\label{cond-WH}
\ee
In fact,
the condition \eqref{cond-WH} is always satisfied
if $F(a) = 0$ and $C(a) \neq 0$.

First, 
consider the possibility that $\rho'(r)$ diverges at $r = a$.
We expand $C(r)$ in the limit $r \to a$ as
\begin{equation}
C(r) = C(a) + 2 \rho_0 (r-a)^n + \cdots \ , 
\label{C-expand-2}
\end{equation}
where $0 < n < 1$ in order for $\rho'$ to diverge at $r=a$, 
Plugging \eqref{sol-F} back to \eqref{Ein-A-uu} or \eqref{Ein-A-uv} 
and expand around $r=a$ by using \eqref{C-expand-2}, 
we obtain 
\begin{align}
 0 &= (r-a)^{n-2} a^2 C(a) \rho_0 n(n-1) 
 \left[C(a) + 2 \kappa a^2 \left(T_{uu}(a) - T_{uv}(a)\right)\right] 
 + \mathcal O\left((r-a)^{2n-2}\right) \ . 
\end{align}
This implies that the condition \eqref{cond-WH} must be satisfied 
even if $\rho'$ diverges as $r\to a$,  
and hence, \eqref{cond-WH} is a necessary condition to have a wormhole 
geometry near $r=a$, independent of whether $\rho'$ diverges or not. 

With the expansion \eqref{C-general-expansion} and \eqref{F2-general-expansion}
for $C(r)$ and $F(r)$,
the absence of horizon ($C(a) \neq 0$) means that 
\be
K = 0 \ .
\ee
The equations for the metric \eqref{metric-general} 
and those for the energy-momentum tensor \eqref{T4vu-general}--\eqref{T4thth-general} remain valid.

Depending on the value of $K'$ and $M$,
the solutions that resemble wormholes are characterized as follows.
\begin{enumerate}
\item
If $K' = 0$,
\bea
\kappa T^{v}{}_{u}(a) &=& G^{v}{}_{u}
= \frac{f_0}{2a c_0} > 0 \ ,
\\
\kappa T^{u}{}_{u}(a) &=& G^{u}{}_{u}
= - \frac{1}{a^2} + \frac{f_0}{2a c_0} 
= - \frac{1}{a^2} + \kappa T^{v}_{u}(a) > - \frac{1}{a^2} \ ,
\\
\kappa T_{\th\th}(a) &=& G_{\th\th}
= \mbox{depends on $M$} \ ,
\\
ds^2 &\simeq& - c_0 dt^2 + \frac{c_0}{f_0^2 (r-a)} dr^2 + a^2 d\Omega^2
\nn \\
&\simeq& - c_0 dt^2 + c_0 dr_*^2 + a^2 d\Omega^2 \ ,
\eea
where 
$\frac{f_0}{2ac_0} \sim {\cal O}\left(\frac{1}{a^2}\right)$
and $r = a + \frac{f_0^2}{4} r_*^{2}$ $(r_* \geq 0)$.
This is a wormhole with the neck at $r_* = 0$.

\item
If $K' > 0$ and $K' < M$, 
\bea
\kappa T^{v}{}_{u}(a) &=& G^{v}{}_{u}
= 0 \ ,
\\
\kappa T^{u}{}_{u}(a) &=& G^{u}{}_{u}
= - \frac{1}{a^2} \ ,
\\
\kappa T_{\th\th}(a) &=& G_{\th\th}
= \mbox{depends on $M$} \ ,
\\
ds^2 
&\simeq& - c_0 dt^2 + c_0 dr_*^2 + a^2 d\Omega^2 \ ,
\eea
where
\begin{equation}
 r = a + \left[\frac{(M-K')f_0}{2M} r_* \right]^{2M/(M-K')} \ . 
\end{equation}
By rewriting 
\begin{equation}
 \frac{2M}{M-K'} = \frac{p}{q}
\end{equation}
where $p$ and $q$ are co-prime integers, 
the geometry has the wormhole structure if $p$ is even,
and $r\geq a$ for arbitrary $r_*$. 
If neither $p$ nor $q$ is even, 
we have $r>0$ for $r_*>0$ and $r<0$ for $r_*<0$. 
If $q$ is even, the above coordinates are well defined only for $r_*>0$. 

\item
If $K' > 0$ and $K'\geq M$, 
\bea
\kappa T^{v}{}_{u}(a) &=& G^{v}{}_{u}
= 0 \ ,
\\
\kappa T^{u}{}_{u}(a) &=& G^{u}{}_{u}
= - \frac{1}{a^2} \ ,
\\
\kappa T_{\th\th}(a) &=& G_{\th\th}
= \mbox{depends on $M$} \ ,
\\
ds^2 
&\simeq& - c_0 dt^2 + c_0 dr_*^2 + a^2 d\Omega^2 \ ,
\eea
where
\be
r = 
\begin{cases}
a + e^{f_0 r_*} & (K' = M) \ ,
\\
r = a + \left[- \frac{(K' - M)f_0}{2M} r_* \right]^{-2M/(K'-M)} & (K' > M) \ .
\end{cases}
\ee
In these cases, the point $r=a$ corresponds to $r_*\to-\infty$. 
The speed of light is $dr_*/dt = 1$,
hence it takes an infinite amount of time (change in $t$)
to reach the point $r = a$ from the viewpoint of a distant observer.

\end{enumerate}

For all the wormhole-like geometries,
the energy-momentum tensor must satisfy the condition \eqref{cond-WH}
and $T^{v}{}_{u}(a) \geq 0$.
($T^{v}{}_{u}(a)$ must be zero or positive for $F(a)=0$.)
If $T_{uu}(r)$ is always positive,
the geometry has neither horizon nor wormhole-like structure.

\section{Conclusion}

In Secs.~\ref{ToyModel}~and~\ref{dilaton-coupled-static},
we considered different models of the vacuum energy-momentum tensor,
and studied its back reaction on the geometry.
We summarize our results as follows.

\begin{enumerate}
\item
The perturbation theory for the Schwarzschild background
breaks down at the horizon
(in the Schwarzschild coordinates)
in the expansion of Newton's constant.

\item
The Schwarzschild metric is modified
in a very small neighborhood of the Schwarzschild radius
($r - a_0 \ll \alpha/a_0$) by the quantum correction
to the energy-momentum tensor.

\item
For the Boulware vacuum, 
there is no horizon
for the model considered in Sec.\ref{ToyModel}.
Instead,
there is a wormhole-like geometry
near the Schwarzschild radius.
For the model considered in Sec.\ref{dilaton-coupled-static},
there may or may not be a horizon,
or a wormhole-like geometry,
depending on the vacuum state.

\item
For the model considered in Sec.\ref{ToyModel},
if there are non-zero energy flows
in the asymptotic region with an appropriate intensity, 
there is a fine-tuned solution with a horizon.
Generic solutions have the wormhole-like geometry
instead of the horizon.

\item
In all cases considered,
the magnitude of the Einstein tensor ($G^u{}_u, G^v{}_u, G^{\th}{}_{\th}$)
is of order ${\cal O}(1/a^2)$ or smaller.

\end{enumerate}

These results are in contradiction with the conventional folklores that
a small quantum correction%
\footnote{Of course,
a classical correction to the energy-momentum tensor would have
exactly the same effect through Einstein's equations.}
would not destroy the horizon,
and that the Boulware vacuum has a diverging (or Planck-scale) 
energy-momentum tensor at the horizon. 
The diverging quantum effects at the horizon in the classical black hole geometries 
imply modification of the saddle point of path integral,  
by the quantum effects.%
By taking the back reaction from the quantum effects into account, 
the geometry is modified at the horizon such that 
the energy-momentum tensor has no divergence, and then, 
the Boulware vacuum gives physical configurations. 

The calculations leading to the results mentioned above
demonstrated a connection between the vacuum energy-momentum tensor
and the near-horizon/wormhole-like geometry.
Hence we explored in Sec.~\ref{General}
this connection for generic energy-momentum tensors,
for solutions with a horizon or a wormhole-like structure.
We summarize the results as follows.

\begin{enumerate}

\item
If $T_{uu}$ (which equals $T_{vv}$) or $T_{uv}$
is non-vanishing around the Schwarzschild radius,
regardless of how small they are,
there can be no horizon.

\item
If $T_{uu}(a)=T_{uv}(a)=0$ and $T^u{}_u(a) > - \frac{1}{\kappa a^2}$, 
the geometry can have the horizon at $r=a$, and must be the Rindler space near the horizon, 
the same as the Schwarzschild black hole. 

\item
If $T_{uu}(a)=T_{uv}(a)=0$, 
$T^u{}_u(a) = - \frac{1}{\kappa a^2}$ and $T_{\th\th}(a) > 0$, 
the geometry can have the horizon at $r=a$, and the near-horizon geometry 
is given by Rindler space or $AdS_2 \times S^2$, 
the same as that of the Schwarzschild black hole or 
the extremal Reissner-Nordstr\"{o}m black hole, for example, respectively.

\item
If $T_{uu} = T_{vv}$ is negative at $r=a$, and $T_{uu}$ and $T_{uv}$ satisfy 
\begin{equation}
 T_{uu}(a) - T_{uv}(a) = - \frac{C(a)}{2\kappa a^2} \ ,
\end{equation}
the geometry cannot have the horizon there, 
but can have the wormhole-like structure, i.e.\ the function 
$r$ can have a local minimum there.

\item
if $T_{uu} = T_{vv}$ is positive around Schwarzschild radius, 
there would be no horizon nor wormhole-like structure. 

\end{enumerate}

In particular, 
the models considered in Secs.~\ref{ToyModel}~and~\ref{dilaton-coupled-static}
demonstrate that the necessary condition for the horizon
(See item 1) 
is not guaranteed as a robust nature of the matter fields. 
Although it is natural that the energy-momentum tensor vanishes in the bulk at the classical level, 
the quantum effects provide non-zero $T_{uu}$ and $T_{vv}$ in general. 
The horizon should be viewed as a rare structure that demands fine-tunning.

The readers may have reservations for some of the assumptions we made,
such as the validity of the Einstein equation,
the spherical symmetry,
or the quantum models used to calculate the vacuum energy-momentum tensor.
Even if all of these assumptions are not reliable,
our work should have raised reasonable doubt against the common opinion
that the back reaction of quantum effects 
can only have negligible effect on the existence of the horizon
\cite{Bardeen:1981zz,Abdolrahimi:2016emo}.
In the examples we studied,
the existence of the horizon is sensitive to the details 
of the energy-momentum tensor.

It will be interesting to extend our analysis to 
the dynamical processes of gravitational collapse. 
In this paper, we have studied static geometries for which
the Killing horizon, event horizon and apparent horizon coincide,
but they could be different in time-dependent geometries.
For a gravitational collapse, 
the initial spacetime is typically the flat spacetime. 
At a later time,
it would approximately be the Unruh vacuum near the Schwarzschild radius instead of the Boulware vacuum. 
(It is not exactly same to the Unruh vacuum since 
 the boundary condition should be imposed at the past horizon for the Unruh vacuum.) 
There would be outgoing energy flux corresponding to Hawking radiation at large $r$,
and the energy-momentum tensor near the surface of the star would also be modified. 
With this correction to $T_{uu}$,
the status of the future horizon can be affected. 
The qualitative nature of the space-time geometry
at a given constant $u$ is expected to resemble
that of the static geometry
(e.g. the wormhole-like structure).
The Killing horizon can be excluded as discussed in Sec.~\ref{General} 
if there is non-zero outgoing energy flow. 
The apparent horizon and event horizon, however, can in principle appear.
Nevertheless,
let us not forget that the expectation of a horizon 
in the conventional model of gravitational collapse
is based on our understanding of the static Schwarzschild solution,
and we have just shown that 
the horizon of the Schwarzschild solution can be easily removed 
by the back reaction of the vacuum energy.
We believe that a better understanding of the static black holes 
would allow us to describe the dynamical black holes more precisely.

For the cases of wormhole-like geometries,
inside the throat (or turning point),
the outgoing null geodesics converge and 
ingoing null geodesics diverge. 
If there is a matter inside the wormhole, 
the structure along the outgoing null geodesics 
is qualitatively same to the conventional model of the black hole evaporation. 
The structure along the ingoing null geodesics, which is 
different from the conventional model, would possibly be modified 
when the time evolution due to the evaporation process is taken into account. 
From the viewpoint of a distant observer,
this scenario is compatible with the conventional model,
although the space-like singularity at $r = 0$
would be replaced by the internal space inside the throat,
that is, a bubble of space-time attached to the outer world through
a throat of 0 or Planckian-scale radius.
More details about this scenario of gravitational collapse will be reported
in a separate publication.

Another scenario of gravitational collapse is described by,
the KMY model \cite{Kawai:2013mda}
(see also \cite{Kawai:2014afa} -- \cite{Kawai:2017txu}),
which are given by exact solutions to the semi-classical Einstein equation \eqref{Einstein-Eq}, 
including the back reaction of Hawking radiation. 
It was shown that 
Hawking radiation is created only when the collapsing shell is still
(marginally) outside the Schwarzschild radius.
If the star is completely evaporated into Hawking radiation within finite time, 
regardless of how long it takes, the apparent horizon would never arise. 
In the KMY model, 
just like our results for the static black hole, 
the horizon is removed due to a modification of the geometry 
within a Planck-scale distance from the Schwarzschild radius 
due to the back reaction of the energy-momentum tensor of the quantum fields. 
While different quantum fields can have different contributions
to the vacuum energy-momentum tensor,
we believe that the general connection between the energy-momentum tensor and the near-horizon geometry
will be important for a comprehensive understanding on the issue of the formation/absence of horizon.
This work is a first step in this direction.

There are other works 
\cite{Gerlach:1976ji,FuzzBall,FuzzBall2,Barcelo:2007yk,Vachaspati:2006ki,Krueger:2008nq,Fayos:2011zza,
Mersini-Houghton,Saini:2015dea,Baccetti} 
that have also proposed the absence of horizon in gravitational collapse
based on different calculations.
However, it might be puzzling to many
how the conventional picture about horizon formation could be wrong.
We find most of the arguments for the formation of horizon neglecting
the vacuum energy's modification to geometry within a Planck scale distance from the Schwarzschild radius.
This paper points out that these approximations are not reliable.

\section*{Acknowledgement}

The authors would like to thank 
Hikaru~Kawai for sharing his original ideas,
and to thank
Jan~de~Boer,
Yuhsi~Chang, Hsin-Chia~Cheng, Yi-Chun~Chin, 
Takeo~Inami, 
Hsien-chung~Kao, Per~Kraus,
Matthias~Neubert, 
Shu-Heng~Shao, 
Masahito~Yamazaki, I-Sheng~Yang, Shu-Jung~Yang and Xi~Yin 
for discussions.
P.M.H. thanks the hospitality of the High Energy Theory Group at Harvard University,
where part of this work was done.
The work is supported in part by
the Ministry of Science and Technology, R.O.C. 
(project no.~104-2112-M-002-003-MY3)
and by National Taiwan University.


\end{document}